\documentstyle[prb,aps,epsfig,twocolumn]{revtex}
\begin{document}
\draft
\flushbottom
\twocolumn[
\hsize\textwidth\columnwidth\hsize\csname @twocolumnfalse\endcsname

\hyphenation{ad-van-ces Kol-mo-go-rov Mo-ser}
\author{R. E. Prange, R. Narevich and Oleg Zaitsev}
\address{Physics Department, University of Maryland, College Park, MD 20742}
\title{Quasiclassical Surface of Section Perturbation Theory}
\maketitle

\begin{abstract}
Perturbation theory, the quasiclassical approximation and the quantum
surface of section method are combined for the first time. This gives a new
solution of the the long standing problem of quantizing the resonances
generically appearing in classical perturbation theory. Our method is
restricted to two dimensions. In that case, however, the results are
simpler, more explicit and more easily expressed visually than the results
of earlier techniques. The method involves expanding the `phase' of the
wavefunction in powers of the {\em square root} of the small parameter. It
gives explicit WKB-like wavefunctions and energies even for certain systems
which classically show hard chaos.
\end{abstract}
]

\narrowtext
\tightenlines

\section{Introduction and summary}

\subsection{Generalities}

The relationship of a quantum system and its classical limit has been a
major theme in physics for nearly a century. Quasiclassical, or WKB
approximations to quantum theory helped elucidate the relation of the full
theory to the classical limit, but this was largely restricted to the
integrable case. Van Vleck\cite{VanVleck} applied this method more
generally, but there was little work going much beyond the integrable case
until 1970 when Gutzwiller\cite{Gutz} introduced his trace formula, [GTF].

Shortly before this, significant advances had been made in classical
mechanics. Notably, Kolmogorov, Arnol'd and Moser proved their famous KAM
theorem\cite{Gutz,LandL,Lazutkin}, which we discuss further below. Related
to this and equally important, new ideas and related numerical studies made 
{\em chaos }an important field of study. In practice, the GTF is applied
only to `hard' chaos systems, where all classical orbits are unstable. A
related trace formula, due to Berry and Tabor\cite{Gutz}, applies to the
integrable case. However, a very large class of systems, the `mixed' chaos
case, has not been so successfully treated by trace formulas.

It should be appreciated that the trace formula, by itself, does not solve
the quantum hard chaos problem. At best, it gives energy levels, not wave
functions. Only a few of the lowest levels are even approximately given,
however. It is mainly convenient for the study of correlations of widely
separated energy levels, which is a short time property. However, it is
certainly true that two decades of efforts to understand the GTF and to
improve it resulted in techniques which did successfully solve the energy
level problem at the WKB level. One of those techniques, due to Bogomolny%
\cite{bogolss}, is the basis of the present work. This method is closely
related to that of Doron and Smilansky\cite{dorsmil}, and more subtly to the
ideas of Berry and Keating\cite{BK} and of Cvitanovi\v c and Eckhardt\cite
{CvitEck}.

\subsection{Perturbation theory}

It is not necessary to emphasize that approximations based on large or small
quantities are important in mathematics. Indeed, KAM theory is a resolution
of problems of classical perturbation theory first uncovered by Poincar\'e.
Attempts have been made, with some success, to {\em quantize }a generically
perturbed classically integrable system. Of course, until KAM the classical
system itself was poorly understood. However, KAM basically found that a
perturbed system generically looks rather like a mixed chaos system, and we
have remarked that quantum mixed chaos is far from being understood.

On the other hand, given an integrable {\em quantum }system, perturbation
theory on it is routinely carried out. This is relatively easy for the case
when $\hbar $ is not `small', that is, when the perturbation is smaller than
the energy level spacing.

\subsection{Two small parameters}

In this paper we are interested in the case where {\em two} small parameters
are present. The first parameter, denoted $\epsilon $, gives the scale of
the difference between an exactly solvable, integrable case, and the
`perturbed' case of interest. The second parameter is the dimensionless
Planck's constant, $\hbar ,$ which gives the scale of the leading order
quantum or wave effects.

The {\em relationship} between these parameters is crucial, and the theory
is exceeding rich as a result. Although this has certainly been understood
for some time, we have not found a very clear discussion in the context of
the formal perturbation-quasiclassical approaches used up to now.

The relationship comes from the well known fact, made manifest by Feynman's
path integral formulation, that quantum effects smear classical phase space
structures over areas of size $h=2\pi \hbar .$ We choose to compare Planck's
constant with the area of the leading phase space structure, call it ${\cal S%
}.$ The {\em dimensionless} Planck's constant is $h/{\cal S}$, which we
continue to call $h,$ i.e. we choose ${\cal S}$ as the unit of action. For
example, in a nearly circular billiard of radius $R,$ ${\cal S}$ could be $%
{\cal S}=pR=\hbar kR$, so $h/{\cal S}=2\pi /kR=\lambda /R.$ Here $p$ is the
momentum, $k$ the wavenumber and $\lambda $ the wavelength of the particle
in the billiard. If the dimensionless Planck is of order unity, only the
gross features of the classical system are reflected in the quantum
properties. But for small $\epsilon $, or in the presence of chaos, there
are typically classical structures on all scales, i.e. on the scale of $%
\epsilon ^\beta ,$ where $\beta $ ranges from zero to infinity. Typically,
the smallest nonvanishing $\beta $ of interest is $\beta =\frac 12.$ If $%
\epsilon ^M/h$ is of order unity, it means that quantum systems do not
reflect the structure at the levels $\epsilon ^\beta $ with $\beta >M$ but 
{\em are} sensitive to structure with $\beta \leq M.$

\subsection{Integrable classical systems}

An integrable classical system has enough constants of the motion that each
classical orbit in $2d$ dimensional phase space lies on a $d$-dimensional
surface in that space. These surfaces turn out to be tori. The $d$ constants
of the motion can be taken to be the {\em action variables, }$I_j,$ $j=1..d,$
and they label the {\em invariant tori. }{\bf From now on, we take} $d=2.$
In these canonical variables, the Hamiltonian $H_0$ depends only on the
actions, i.e. $H_0(I_1,I_2)=E$, where $E$ is the energy. The conjugate {\em %
angle variables}, $\theta _1,\theta _2$, satisfy Hamilton's equations $\dot 
\theta _j=\partial H_0/\partial I_j\equiv \omega _j=\omega _j(I_1,I_2).$ The
angle variables fix the point on a given invariant torus. {\bf We shall also
assume}, to avoid certain complications, a principle of `sufficient
nonlinearity', which posits that the winding number $\omega _1/\omega _2$ is
not too constant as $I_1$ and $I_2$ are varied at fixed $E.$ We also often
use two other constants equivalent to the $I$'s, usually $E$ and some other
convenient variable, such as the winding number.

A common way of describing the perturbation is to give a Hamiltonian 
\begin{equation}
H=H_0(I_1,I_2)+\epsilon H_1(I_1,I_2,\theta _1,\theta _2)  \label{H}
\end{equation}
where the perturbation is periodic in the angles. It is well known in this
field, however, that other formulations are often more convenient.

\subsection{Poincar\'e surface of section}

A Poincar\'e surface of section, SS, is a phase space surface through which
all tori of interest pass, once and only once. In action angle variables, or
their equivalent, we usually take the surface $\theta _2=0,$ $E$ fixed. Fig.
1 illustrates such a surface. A given orbit crosses the SS each time $\theta
_2$ is a multiple of $2\pi $. Given $\theta _1^{\prime }$ and $I_1^{\prime }$
when $\theta _2=0,$ the dynamics predicts a new $(I,\theta )={\cal T}%
(I^{\prime },\theta ^{\prime })$ at the next intersection of the orbit with
the SS. (From now on, we call the variables on the SS simply $I,\theta ,$
without subscripts.) ${\cal T}$ is called the {\em surface of section map}
and of course it depends on $E.$ For an integrable system, $I=I^{\prime },$ $%
\theta =\theta ^{\prime }+2\pi \omega _1/\omega _2.$ For a non integrable
system, $I$ is not constant.

The surface of section is usually displayed in a graph of $I$ {\em vs} $%
\theta $, for $0\leq \theta <2\pi ,$ and with points $0,\,2\pi $ identified.
The intersection of the invariant torus $I=const$ with the surface of
section, is the horizontal line $I=const$ on this graph.

The surface of section map is conveniently given by a {\em generating
function}. This is an action, $S(\theta ,\theta ^{\prime })$ such that $%
I=\partial S(\theta ,\theta ^{\prime })/\partial \theta $, $I^{\prime
}=-\partial S(\theta ,\theta ^{\prime })/\partial \theta ^{\prime }.$ In the
integrable case, $S(\theta ,\theta ^{\prime })=S_0(\theta -\theta ^{\prime
})=(\theta -\theta ^{\prime })I_1+2\pi I_2$, where the actions are regarded
as functions of $\theta -\theta ^{\prime }=2\pi \omega _1/\omega _2$ and $E.$
We note that $S$ is just the integral $\int pdq$ along the trajectory from
one surface of section crossing to the next.

\begin{figure}[tbp]
{\hspace*{0.7cm}\psfig{figure=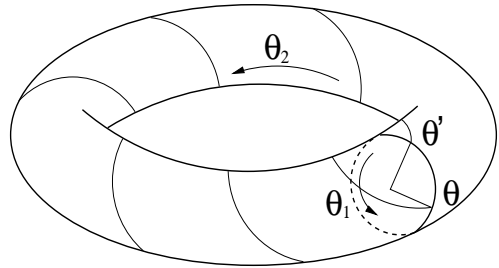,height=6.5cm,width=3.5cm,angle=270}} {%
\vspace*{.13in}}
\caption[ty]{Schematic of the definition of surface of section $%
\theta_2=const$. An orbit starts at $\theta_1=\theta ^ \prime$ in the SS,
goes around the hole once, arriving at $\theta_1 = \theta$. It advances by
an angle $\theta - \theta^\prime = 2\pi \omega _1/\omega _2$ on the SS.}
\label{fig:1}
\end{figure}

Under perturbation, $S$ becomes 
\begin{equation}
S(\theta ,\theta ^{\prime })=S_0(\theta -\theta ^{\prime })+\epsilon
S_1(\theta ,\theta ^{\prime })+O(\epsilon ^2).  \label{S}
\end{equation}
There is no difficulty, except tedium, in using perturbation methods to find 
$S_1$ given say, $H_1$, since the calculation involves only short orbits.
That is to say, the fundamental difficulties with classical perturbation
theory lie in understanding the long time behavior.

\subsection{Quantum surface of section method}

Our theory is expressed in terms of $S.$ Bogomolny\cite{bogolss} has shown
(without requiring $\epsilon $ small) that the surface of section transfer
operator or kernel 
\begin{equation}
T(\theta ,\theta ^{\prime })=\left( \frac 1{2\pi i\hbar }\frac{\partial
^2S(\theta ,\theta ^{\prime })}{\partial \theta \partial \theta ^{\prime }}%
\right) ^{\frac 12}\exp \left( \frac i\hbar S(\theta ,\theta ^{\prime
})\right)  \label{T}
\end{equation}
can be used to find the energy levels of a system, in quasiclassical
approximation, although generally the solution is not very explicit. We
mention some of the main results below. Bogomolny's method is by no means
restricted to action angle variables or any particular surface of section.
There is a large literature\cite{TLit,PranFred} exploiting and verifying
this technique. We call this the {\em quantum surface of section method. }We
shall find that using Eq.(\ref{S}) in Eq.(\ref{T}) and exploiting the small
parameter $\epsilon $, allows a rather complete and explicit solution of the
quantum problem.

\subsection{KAM theory}

We summarize KAM theory as follows. A Hamiltonian $H=H_0+\epsilon H_1$ is
assumed, where $H_1$ is a suitably smooth and bounded function of $%
I_j,\theta _j,$ and is periodic in the angle variables. It is found that,
generically, the original invariant tori which are {\em rational}$,$ or
equivalently {\em resonant, }are destroyed, as well as those in their
immediate vicinity. A rational torus is one for which the winding number $%
\omega _1/\omega _2$ is a rational, $p/q.\,$ Any orbit on such a torus whose 
$\theta _2$ increases by $2\pi q$ will have $\theta _1$ increase by $2\pi p$
: it is thus a {\em periodic }orbit. The Poincar\'e-Birkhoff theorem\cite
{LandL} says that only a few, often just two, such orbits of the original
continuum on the torus, survive the perturbation. One surviving orbit is
stable, the other unstable.

Within a width in action, $\sqrt{\epsilon }I_{pq}$ about the rational torus,
the neighboring tori, both rational and irrational, are destroyed or
modified strongly. Here $I_{pq\text{ }}$ is a characteristic classical scale
of action associated with the $pq$ torus.

The rational tori are of course dense. However, the series $\sum_{pq}I_{pq}\ 
$converges to a finite action, if the perturbation is smooth enough, so the
total volume in which the original invariant tori are destroyed is a small
but finite fraction of phase space, proportional to $\sqrt{\epsilon ,}$.

We may `cut out' this phase space volume near the original rational tori.
What is left is called a {\em KAM set}\cite{Lazutkin}{\em . }The KAM set is
a Cantor set, which however occupies most of the volume of phase space. The
tori in the KAM set are invariant and only slightly modified under
perturbation. This is the main result of KAM theory.

The destroyed tori are largely replaced by {\em new }invariant tori. These
have a new topology, namely the topology of tubes with the surviving {\em %
stable }periodic orbit(s) at the center(s). In the surface of section
representation of Fig. 2, these look like islands.

\begin{figure}[tbp]
{\hspace*{-0.2cm}\psfig{figure=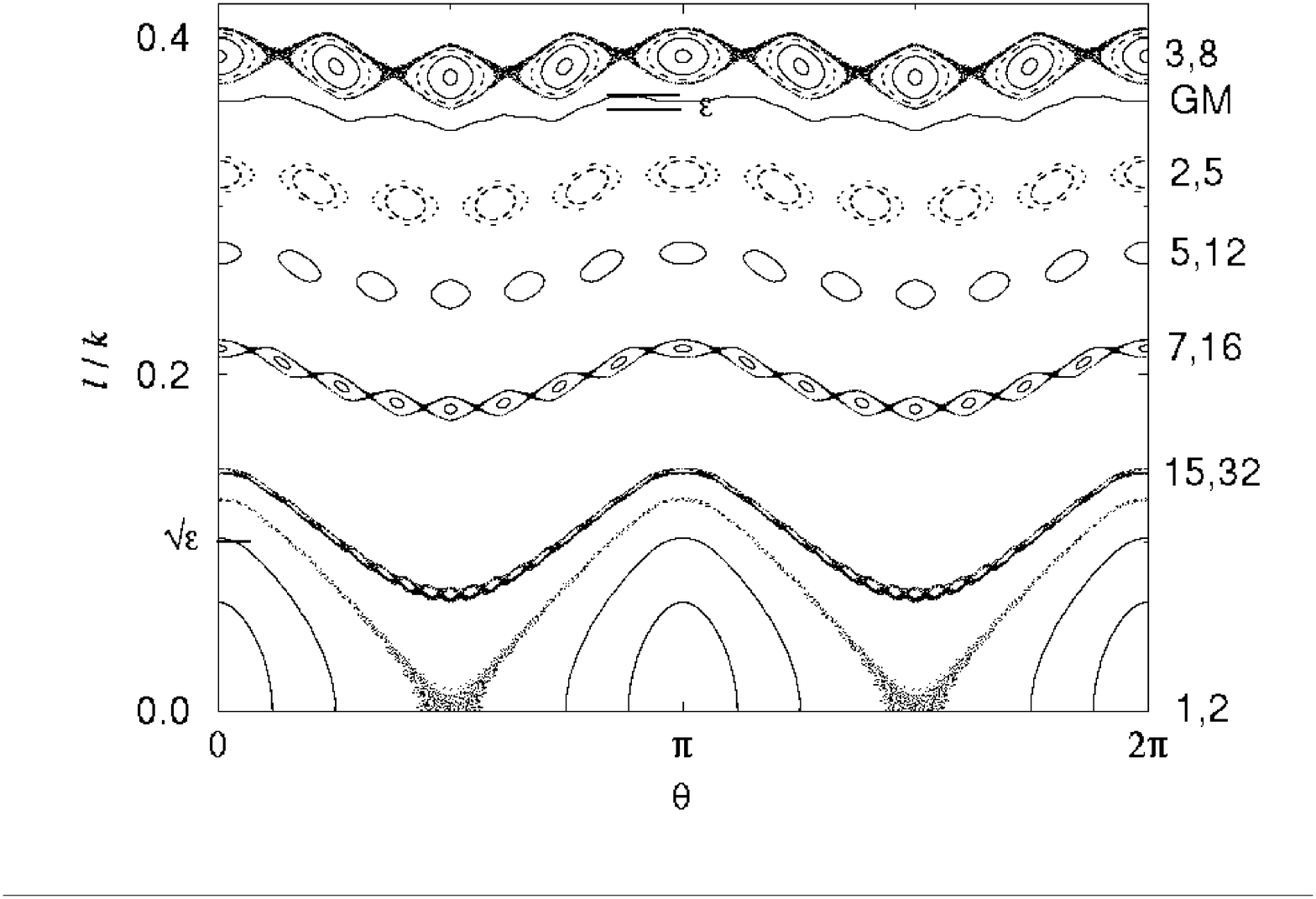,height=6cm,width=8.5cm,angle=0}} {%
\vspace*{.13in}}
\caption[ty]{Surface of section invariant tori and separatrices, showing the
scale $\protect\sqrt{ \epsilon }$ of resonances, the scale $\epsilon $ of
the `golden mean' torus in the KAM set, and secondary resonances, all for $%
\epsilon=0.01$, $\eta =0.19$.}
\label{fig:2}
\end{figure}

Separating these tubes from one another and from the KAM set is a separatrix
region, consisting of the stable and unstable manifolds of the remaining 
{\em unstable }periodic orbit. Because these manifolds intersect
transversely at a homoclinic point, although at a very small angle, this
separatrix region becomes a chaotic, homoclinic tangle, also referred to as
a stochastic region. Outside the separatrix, there are invariant tori,
perturbed on the scale of $\sqrt{\epsilon }I_{pq}$ which `gradually' go into
the KAM set.

Fig. 2 shows such intersections of invariant tori with the surface of
section, which we call invariant loops. These represent tori in the KAM set,
new invariant tori and separatrices.

The new invariant tori can also be classified as rational or irrational, and
the whole process must be repeated, {\em ad infinitum, }giving rise to
secondary resonances, and so on. However, the original perturbation scale $%
\epsilon ,$ is replaced for the secondary resonances\cite{LandL} by
something like $\epsilon ^{1/\epsilon }$, which is smaller than any power of 
$\epsilon .$

The scale of the stochastic region is also very small\cite{Lazutkin}, namely
of order $\exp (-1/\sqrt{\epsilon }),$ again beyond any ordinary
perturbation theory. These phenomena are represented schematically in Fig. 2.

KAM theory thus shows that after perturbation phase space consists of an
infinite number of parts.

\begin{enumerate}
\item  A large part, the KAM set, with a Cantor set structure, of relative
phase space volume of order $1-\sqrt{\epsilon }$ but weakly and smoothly
perturbed on the scale of $\epsilon .\,$

\item  A relatively big part, consisting of new invariant tori near the
original rational tori, which usually has a scale of $\sqrt{\epsilon },$ but
made up of an infinite number of pieces of size $\sqrt{\epsilon }I_{pq}.$
Remark that $\sqrt{\epsilon }$ is bigger than any power of $\epsilon ,$
which is one reason that perturbation theory in $\epsilon $ must fail.

\item  Conceptually important but smaller-than-any-power-of-$\epsilon $
stochastic regions which also are near the original resonant tori.

\item  The still smaller secondary resonances and stochastic regions, near
the new rational tori.

\item  The smaller yet tertiary resonances and stochastic regions, ad
infinitum.
\end{enumerate}

A final remark for this section. The appearance of the square root $\sqrt{%
\epsilon \text{ }}$ is generic in the sense that it occurs when the second
derivative $S_0^{\prime \prime }(\theta )$ does not vanish at a value of $%
\Theta _{pq}=2\pi p/q$ explained further below. It is easy to construct
examples for which this fails. Our method can be generalized to deal with
such cases.

\subsection{Results}

Most of these results have been announced\cite{pnz1}. We are able to find
explicitly all energy levels and `wave functions' to leading order in $\hbar
.$ By wave functions, we mean eigenstates of the $T$ operator, which are
closely related to the usual wave functions. The equation we solve is 
\begin{equation}
T\psi =\psi   \label{Tpsi}
\end{equation}
This equation can be solved only if $E,$ on which $T$ depends
parametrically, is on the spectrum or more precisely is on the WKB
approximation to the spectrum.

There generally exists an exact operator or kernel $K$ to which $T$ is a
quasiclassical approximation, for which this procedure gives exact answers%
\cite{PranFred}. There is a Fredholm integral equation over the SS 
\begin{equation}
\mu (\theta ^{\prime },{\bf r})=\nu (\theta ^{\prime },{\bf r)+}\int d\theta
K(\theta ^{\prime },\theta ;E)\mu (\theta ,{\bf r).}  \label{Fred}
\end{equation}
and the Green's function $G({\bf r}^{\prime }{\bf ,r;}E{\bf )}$ can be
computed from the solution $\mu .$ Eq.(\ref{Tpsi}) is the homogeneous
version of Eq.(\ref{Fred}) with $K$ approximated by $T.$ For billiards, with
the boundary as SS, Eq.(\ref{Fred}) is the equation for the boundary
integral method, and $\psi $ is the normal derivative of the usual
wavefunction $\Psi ({\bf r)}$ on the boundary.

There is a known procedure to find $\Psi $ given $\psi .$ However, $\psi $
contains most of the information desired rather directly, and it is not
necessary as a rule to find $\Psi .$ In fact, there are many ways available
to represent a complex function of two variables, and $\psi \,$ is a useful
one even in nonperturbative contexts.

We can describe what we have done as the solution of Eq.(\ref{Tpsi}) in WKB
approximation for the wave functions and energy levels as a power series in $%
\sqrt{\epsilon },$ or in $\epsilon ,$ depending on the phase space region.
We also have a criterion for whether the series should be $\sqrt{\epsilon },$
or $\epsilon .$ Thus, in principle, we find explicit solutions for those
quantities which are not smaller than any power of $\epsilon .$ To be a bit
more accurate, we find explicit solutions for the leading $\sqrt{\epsilon }$
and first few powers, and provide a procedure which rapidly becomes tedious,
for the higher orders. In this paper we will concentrate on the leading
order or two, and not on issues of the breakdown of the series, etc.

A second interpretation is in terms of quantum perturbation theory. We do a
quantum perturbation theory about an initial state $\psi =\exp (il\theta ).$
However, unless $\sqrt{\epsilon }/\hbar <<1$, this is not an ordinary
perturbation theory, but rather a {\em degenerate }perturbation theory, in
which of order $\sqrt{\epsilon }/\hbar $ unperturbed states are mixed, in
the resonant case, or $\epsilon /\hbar $ states in the KAM set case. This
can be seen in Fig. 2, when it is realized that unperturbed states are
represented by action values spaced $\hbar $ apart. So, we can effectively
carry out the diagonalization of the matrix mixing these states in this case.

The fact that a form of degenerate perturbation theory is needed was
apparently first pointed out by Shuryak\cite{Shuryak}, in the case of
resonances. Thus, if $\sqrt{\epsilon }I_{pq}<<\hbar $ ordinary quantum
perturbation theory works well. Above the `Shuryak border', $\sqrt{\epsilon }%
I_{pq}\geq \hbar $, ordinary perturbation theory breaks down as a number $%
\sqrt{\epsilon }I_{pq}/\hbar $ unperturbed quantum states are strongly mixed
by the perturbation. However, even in the case of the KAM set, if $\epsilon
/\hbar $ is large, many unperturbed levels must be mixed to obtain a
solution.

A third interpretation of what we do is that we are able to find explicit
approximate formulas for the invariant loops. These loops have the topology
of circles and are invariant under the surface of section map. The loops are
found as a power series in $\epsilon $ for tori in the KAM set. We find
explicit formulas for the new, resonant, invariant loops as a power series
in $\sqrt{\epsilon }$. We find explicit formulas for the separatrices in
this way also.

The loops can be quantized, in essence by EKB theory, and the formulation
then gives the other quantization of the motion away from the SS.

We also have expressions for the scale of the hole around a resonant torus, $%
I_{pq}.$ Having these formulas, we can find the quasiclassical quantization
conditions and the wave functions.

The series in $\epsilon $ or $\sqrt{\epsilon \text{ }}$ that we find are not
expected to converge, but are probably asymptotic. If they converged, we
would have found exact new constants of the motion, which cannot exist.

We, however, do {\em not} find any stochasticity, or any secondary resonance
phenomena, since these are smaller than any power of $\epsilon .$ This has
the unfortunate implication that our work does not shed much light on the
quantization of mixed chaos. It would indeed be surprising, however, if the
quantized wave functions of extended stochastic regions had simple analytic
expressions. In fact, mixed chaos consists of three regions, a KAM like
region with invariant tori, a region presumably strongly chaotic and
treatable by the GTF and its improvements, and a transition region between
the two. It is the transition region which has resisted attempts to quantize
it. There is some work\cite{transition} bearing on this transition region
which exploits the fractal nature of the region.

Our results are simple and familiar on the one hand, but rather complex on
the other. That is, we obtain results of a type familiar from standard WKB
theory. But, the results are complex in the sense that the results of
classical KAM theory are complex. Various cases are possible depending on
the choice of system. Since the quantum theory must follow the classical up
to a certain level of approximation, depending on $\hbar ,$ the quantum
results must also be complex in this sense.

\subsection{Other work on combined quasiclassical and perturbation theory}

Combined QCA and PT has been used and studied extensively. The general
scheme is to find approximations for the invariant tori and then use EKB
methods to quantize them. In this sense our approach is a variation of this
already existing work. 

The difference is that we need to find only the invariant loops in the
surface of section, and not the whole torus to which an invariant loop
corresponds. This has the usual advantages of using the surface of section.
It reduces the dimension, thus simplifying calculations and making it easy
to visualize the results. Our results are rather simple to derive and write
down. The older approach, in practice makes approximations for the invariant
tori which are completely numerical or involves expressions with many terms
requiring computer algebra.

An important technique is that of the Birkhoff-Gustavson normal form\cite
{BG,Eckhardt}. This is closely related to the Darling-Dennison form\cite{DD}%
. In this case, the starting point is two or more harmonic oscillator
Hamiltonians coupled by terms consisting of polynomials in the displacements
and momenta. Successive canonical transformations eliminate the coupling
terms, order by order, in favor of a normal form Hamiltonian, which
basically consists of products of powers of the original harmonic
oscillators. Ozorio de Almeida\cite{ozo} has used this to obtain results
somewhat similar to ours. 

This method is particularly relevant in treating vibrational modes of small
molecules. Our method needs modification for this case, as the unperturbed
winding number is constant. On the other hand, it's not clear whether
Birkhoff-Gustavson can be generalized to deal with weakly distorted
billiards, for example.

In practice, Birkhoff-Gustavson relies on computer algebra and numerical
evaluations. Although discussion of the convergence of the series is given%
\cite{Eckhardt}, mention is not made of the relationship of the small
parameter to $\hbar .$

Another approach is to propagate numerically an orbit which lies on an
invariant torus\cite{Ezra2}. Fourier methods then allow a determination of
the fundamental frequencies, the $\omega _i$'s and then the torus itself.
This method is basically numerical, but it is quite efficient.

Although the SS technique is often used to display results, it is seldom
used as a calculational tool. The one exception we have found is the work of
Noid and Marcus\cite{NandM}. Indeed, they used two surfaces of section, SS1
and SS2 which cut the invariant tori in topologically different ways. By
numerical propagation of an orbit, they found the invariant loops $l_{1\text{
}}$ and $l_2$ on these two SS. EBK quantization makes the area of the loops
take the form $area_1=\hbar (n_1+\frac 12),$ $area_2=\hbar (n_2+\frac 12)$,
assuming the  usual $\frac 12$ for librational states. In two dimensions,
the SS, and thus in effect the orbits, depend on two parameters, one of
which is taken as the energy. These parameters are varied until the areas of
the loops have the above relation to integers. Noid and Marcus obtained
numerically good results for the energy levels. Use of SS's gave some
advantage computationally. The method does not rely on a small parameter$,$
but only on the existence of sufficiently simple invariant loops.

A different kind of effort\cite{Ezra1,Ozorio,OandH,Ullmo} studies the
transition away from the Berry-Tabor trace formula, valid for completely
integrable systems at $\epsilon =0,$ and towards the Gutzwiller trace
formula, usually considered only for hard chaos. The Shuryak border and the $%
\sqrt{\epsilon }$ are not mentioned in this work, and an expansion is made
in $\epsilon .$ This seems peculiar, at first sight, since the trace
formulas are purported to give the density of states.

In fact, to get energy levels from the trace formula, a resummation of the
divergent series must be carried out. In other words, the contribution of
very long orbits must be somehow included. One method for doing this uses
Bogomolny's $T$ and is basically equivalent to Eq.(\ref{Tpsi}). The trace
formula without resummation is directly useful if an appropriate energy
average is made, as it predicts energy-level correlations over a correlation
distance long compared with the mean level spacing. This energy average is
equivalent to looking at short time orbits, for which ordinary perturbation
expansions in powers of $\epsilon $ are adequate. The work based on the
trace formula {\em \ }restricts considerations to this case. We can readily
reproduce the cited results and the trace formulas above by use of the
transfer operator.

There are also rigorous mathematical results in this field\cite{Lazutkin}.
We are not sure how to make a simple statement of what has been rigorously
proved. It seems that Lazutkin has succeeded in quantizing the KAM set at
the EKB level, and has shown that the number of levels found is correct. The
wave functions are also correct, up to the possibility discussed below of
accidental degeneracy.

\subsection{Applications}

There are a number of papers\cite{Stone,BCL,FS,Vall} in the recent
literature which have uncovered interesting phenomena in this perturbative
case. Indeed, it was this literature which led us to develop this theory.
Some of this work has practical applications in the construction of tiny
lasers with almost circular resonant cavities\cite{Stone}.

The other work is more concerned with localization theory. The existence of
a rather complete and analytic solution was not suspected, so statistical
studies of localization and energy level statistics were undertaken with
supporting heuristic arguments.

In these papers, the basic object is a weakly deformed circular billiard,
which can be described by a boundary in polar coordinates $R(\theta
)=R_0+\epsilon \Delta R(\theta ).$ In these billiards, the role of $\hbar $
is played by $k^{-1}$, the inverse dimensionless wavenumber or square root
of the energy. Fig. 3 shows such a billiard together with some
characteristic lengths associated with the perturbation theory. The older
literature\cite{MF} treats such boundary perturbations in more ordinary
quantum perturbation theory, valid only for $k\sqrt{\epsilon }<<1.$

The case $\Delta R=\cos 2\theta $ \cite{Stone,Vall} has some special
features in that a nearly circular elliptical billiard has this as its
leading order term, and such a billiard is integrable. Higher order
corrections are quite interesting since they are responsible for island
chains, etc. That is, this case can be thought of as an integrable
elliptical billiard with leading perturbation proportional to $\epsilon ^2.$
We reserve a more detailed analysis to future publication, however.

\subsection{Beyond KAM theory}

KAM theory comes in several flavors, depending on the smoothness of the
perturbation. Examples where KAM does not apply have also been of recent
interest.

A nearly circular Bunimovich stadium billiard\cite{BCL,Borgo,CP} has $\Delta
R\simeq \left| \sin \theta \right| -2/\pi .$ This gives the leading order in 
$\epsilon $ deviation from a circle. We call it the nearly circular stadium
billiard. The constant $2/\pi $ is subtracted to make the average
perturbation vanish, or equivalently, to make the perturbed billiard have
the same area as the unperturbed circle. We shall usually drop it for
brevity, however. For comparison purposes we shall generalize it to $\Delta
R=\sqrt{\sin ^2\theta +\eta ^2}$ to give a billiard with a similar shape,
but with a smoothing parameter $\eta $. We call this the `nearly stadium'
billiard.

It is rigorously known\cite{BCL}, for $\eta =0,$ that because of the lack of
second derivative, the long time behavior is diffusive in angular momentum
space, and there are no invariant tori at all. If $\eta $ is finite but
small, say $\eta <<\sqrt{\epsilon },\,$ similar diffusive results are
obtained numerically. A somewhat related case is to take $\Delta R$ as a
`random' finite Fourier series constrained to have $M$ maxima and minima\cite
{FS}. There is a third, large, parameter $M$ or $1/\eta $, whose
relationship with $k$ and $\epsilon $ must be specified. The possibility of
large higher derivatives of the perturbation can modify or invalidate KAM
theory as well as stationary phase methods.

The general interest in the quantum version of these non KAM cases has been
to find changes of wave function localization behavior as a function of the
parameters, and to interpret these regimes as belonging to different types
of dynamical localization. At some large enough $k$, the large derivatives
in these cases start to play a role, and stationary phase arguments and
perturbation theory breaks down. The contribution of our paper is for values
of $k$ less than this characteristic magnitude. A detailed study of this
breakdown is underway\cite{Zait1}.

However, {\em to some order} in $\sqrt{\epsilon },$ for the resonances,
approximate invariant tori continue to exist, and we can find them and
quantize them. If these tori break down at order $r$, then for $k\epsilon
^{r/2}<<1,$ the quantum system acts as if the tori exist. For larger $k$,
quantum mechanics can `see' the dissolution of the tori, and our theory no
longer applies. This transition and its interpretation is the theme of
several papers\cite{BCL,FS,Borgo,CP}.

This can also be interpreted as the fact that quantum mechanics and its
limiting classical mechanics may have quite different long time behavior.
The two approximately coincide up to a time $\tau _\hbar $, a time usually
interpreted as $\hbar /\Delta $ where $\Delta $ is the mean level spacing of
the quantum levels which are appreciable in the initial wave packet. Beyond
this time, the classical behavior is irrelevant to the understanding of the
quantum behavior. Thus, the long time classical behavior, which is the focus
of most classical theory, can be irrelevant unless it also describes the
short time behavior, $t<\tau _\hbar .$ Quantum mechanics does not `care' if
invariant tori are indeed invariant. It only cares that the classical orbit
stays close enough to a torus for long enough that phase interference
effects can establish the quantum states.

A buzz word used in this context\cite{Borgo,CP} is `cantorus', which is a
Cantor set invariant phase space structure replacing an invariant KAM torus
which has disappeared as a parameter is changed. Cantori\cite{Cantori} can
`trap' for a while classical orbits, thus influencing the short time
behavior. Although this concept is well established in certain contexts\cite
{RepC}, it is not clear to us exactly where the cantori are and where they
are not in the present case. In any case the relevant short time behavior
can be described much more simply and cogently than by invoking cantori. The 
{\em invariance} of a cantorus is what makes it interesting and difficult in
the classical context, but this invariance is of no relevance for quantum
systems.

There are also systems which have no explicit small parameter. Nevertheless,
there can be classes of eigenstates which can be obtained by our technique.
The best known of these are the whispering gallery modes in smooth convex
billiards and the quantized `bouncing ball' orbits in the stadium billiard%
\cite{Tanner}. There are also special modes found in work\cite{ABGOP} on a
ray splitting billiard, and this was the first application of the methods of
this paper. To support a quantum state that we can describe in our theory,
there have to be many similar orbits, e.g. nearly nonisolated orbits similar
to the orbits on a KAM torus. Basically, a small parameter appears as the
solution is found. The potentials found below (which are of order $k\epsilon 
$) are in this case replaced by a big potential, but only `low energy'
solutions are accurately found.

\section{Outline of the method}

\subsection{Specific model system: Perturbed circle billiard}

For a perturbed integrable system, it is possible to choose coordinates and
a surface of section so that the $T$ operator, Eq.(\ref{T}), is based on an
action $S$, of Eq.(\ref{S}). To be explicit, we consider a perturbed
circular billiard.

In quantum language, we take units $R_0=1,$ $\hbar =1,$ particle mass = $1/2$%
, so $k=\sqrt{E}$ is the dimensionless wavenumber, equivalent to $1/\hbar $.
We take the billiard boundary $\partial B$ as SS. Then Bogomolny's unitary
operator is\cite{bogolss} 
\begin{equation}
T(\theta ,\theta ^{\prime },k)=-\left( \frac k{2\pi i}\frac{\partial
^2L(\theta ,\theta ^{\prime })}{\partial \theta \partial \theta ^{\prime }}%
\right) ^{\frac 12}\exp \left( ikL\left( \theta ,\theta ^{\prime }\right)
\right)  \label{TBill}
\end{equation}
where $L$ is the chord distance between two points, specified by polar
angles, on $\partial B$. Expanding, 
\begin{eqnarray}
kL(\theta ,\theta ^{\prime }) &=&2k\left| \sin \frac{\theta -\theta ^{\prime
}}2\right| \left( 1+\epsilon \frac{\Delta R(\theta )+\Delta R(\theta
^{\prime })}2\right) +\ldots  \nonumber  \label{LT} \\
\ &=&k(L_0+\epsilon L_2+\ldots .)  \label{LT}
\end{eqnarray}

We assume that $k$ is large, and $\epsilon $ is small. We introduce $b=\sqrt{%
\epsilon }$. The relation between $k$ and $b$ is specified by an integer $M$%
, such that nominally $kb^M\sim 1$, and $kb^{M-1}>>1$ while $kb^{M+1}<<1.$
In {\em phases} that are expanded in powers of $b$, we must keep terms up to 
$kb^M,$ even though they are much smaller than other contributions to the
phase. This is because such phases give changes of order unity to the
wavefunction as the angle is varied. In this paper we shall usually consider 
$M=2,$ i.e., we shall consider $kb^3<<1.$

In prefactors, we may keep only the leading order terms and succeed in
making only fractionally small errors. In finding the prefactor of the
wavefunction we shall need to keep terms of order unity, which we may regard
as being of the same order as $kb^M.$

This enables us to make a convenient, but not strictly necessary,
approximation to $L_{2\text{ }}$above. We will later find that only values
of $\theta -\theta ^{\prime }$, near a fixed angle $\Theta _l$, depending on
the angular momentum range considered, are important. In $L_2$, we replace $%
\sin \frac 12(\theta -\theta ^{\prime })$ by the constant $d=\left| \sin 
\frac 12\Theta _l\right| ,$ an approximation accurate if $kb^3$ is small.

\subsection{Bogomolny Integral Equation}

According to Bogomolny, the energy levels of the system are given in QCA\cite
{bogolss} by solutions $E=E_a=k_a^2$ of 
\begin{equation}
D(E)=\det (1-T(E))=0.  \label{FredDet}
\end{equation}
This equation is usually approached in one of three ways.

\begin{enumerate}
\item  The imaginary part of the logarithmic derivative of $D$, $d\ln (D)/dE$
can be expanded in traces of powers of $T,$ which yields, for example, the
Gutzwiller trace formula, if all periodic orbits are isolated and unstable.
If $T$ is integrable, this gives the Berry-Tabor result, and in the present
case, it gives the perturbed Berry-Tabor results mentioned above\cite
{Ezra1,Ozorio,OandH,Ullmo} The original derivation of these results is more
difficult, and it had the defect that the sum was mathematically rather ill
defined, since it was not absolutely convergent. The organization of the
series by the $T$ operator derivation groups together orbits coming from the
same power of $T$ and at least yields a series which either converges or
diverges.

\item  The Fredholm determinant $D$ may be expanded by the rules of Fredholm
theory\cite{bogolss,PranFred}, giving an absolutely convergent expression,
which in quasiclassical approximation is a finite sum. An important
improvement uses the unitarity of $T$ to make each term in this sum real.
The same traces of powers of $T$ and periodic orbits appear as in the trace
formulas, but organized into `composite orbits'. This is the main result of
`resummation' of the GTF\cite{bogolss,BK,PranFred}.

\item  The kernel $T$ may be represented by a discrete matrix, and
numerically diagonalized. Although this method is disdained by those
committed to periodic orbit theory, it gives very good results.
\end{enumerate}

Our seemingly more difficult technique finds wavefunctions $\psi $
satisfying 
\begin{equation}
\psi (\theta )=\int d\theta ^{\prime }T(\theta ,\theta ^{\prime };k)\psi
(\theta ^{\prime }),  \label{psi}
\end{equation}
which can be done only for $k=k_a.$ Our method is tractable only if the
orbits are nearly nonisolated, as is the case for a perturbed integral
system.

This problem is naturally generalized to 
\begin{equation}
T\psi =e^{i\omega }\psi  \label{psiMap0}
\end{equation}
which can be solved for all $k,$ and $k$ is regarded as a parameter. This
allows a study of perturbed quantum maps in addition to two dimensional
autonomous systems. The phase $\omega $ will be a function of $k$ and
usually, it is rather easy to solve the equation $\omega (k)=2\pi n$ which
gives the $k$ values such that a solution of Eq.(\ref{psi}) exists.

\subsection{Simplest case}

For pedagogical reasons we start with a case special in two respects. First,
we assume that the states are entirely made up of `low' angular momenta,
which means that $\psi $ is slowly varying compared with the variation of
the $T$ operator. From a classical perspective, the relevant orbits will
pass fairly close to the center of the circle. Straight-line orbits of this
class, for the billiard of Fig. 3, pass through a shaded region at the
billiard center.

\begin{figure}[tbp]
{\hspace*{1.cm}\psfig{figure=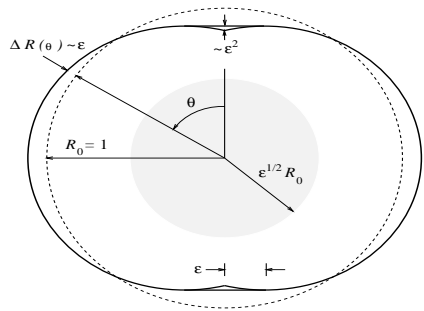,height=5.5cm,width=4cm,angle=270}} {%
\vspace*{.13in}}
\caption[ty]{A nearly circular, nearly stadium billiard $R(\theta) = R_0 +
\Delta R (\theta)$ with $\epsilon = 0.3$. The straight line segments of a
stadium have a length of $2\epsilon$. Its area is equal to the area of the
dashed circle $R_0 = 1$. If approximated to the first order in $\epsilon$, $%
\Delta R(\theta)$ has cusps at $\theta = 0, \pi$. Orbits that cross the
shaded area are affected by $l=0$ resonance. Several relevant lengths are
shown.}
\label{fig:3}
\end{figure}

Second, we assume that $k\epsilon <<1.$ This means that the wavelength is
long compared to the shift of the boundary $\epsilon \Delta R.$ Contrary to
intuition, as we mentioned earlier, this condition does not suffice to make
perturbation results trivial. For that, it is necessary that $k\sqrt{%
\epsilon }<<1,$

With this condition, we may expand the Eq.(\ref{TBill}) as 
\begin{eqnarray}
T(\theta ,\theta ^{\prime }) &\simeq &-\sqrt{\frac k{2\pi i}(1+\ldots )}\exp
(2ik-\frac{ik}4\delta \theta ^2+\ldots )\times  \label{TExpand} \\
&&\times \left[ 1+ik\epsilon V(\theta )+\ldots \right]
\end{eqnarray}
where $\delta \theta =\theta ^{\prime }-\theta -\pi $ is small, i.e. $\theta
^{\prime }\simeq \theta +\pi .$ [We arrange things, to simplify the
notation, so that $\theta ^{\prime }\simeq \theta +\pi $ rather than $\theta
^{\prime }=\theta -\pi .$] Here $V(\theta )=L_2(\theta ,\theta +\pi )=\Delta
R(\theta )+\Delta R(\theta +\pi ).$ We then expand 
\begin{equation}
\psi (\theta ^{\prime })=\psi (\theta +\pi )+\delta \theta \psi ^{\prime
}(\theta +\pi )+\frac 12\delta \theta ^2\psi ^{\prime \prime }(\theta +\pi
)+\ldots  \label{psiExpand}
\end{equation}
We also expand 
\begin{equation}
e^{i\omega (k)}=e^{i\left( 2k+\frac 12\pi +\omega _0\right) }(1+ik\epsilon
E_m+\ldots )  \label{omegaExpand}
\end{equation}
where $E_m$ is a shift of eigenphase to be determined and $\omega _0$ is
defined below. The $\frac 12\pi $ comes from the prefactor and Gaussian
integral over $\delta \theta .$ Using expressions (\ref{TExpand}) , (\ref
{psiExpand}) and (\ref{omegaExpand}) in Eq.(\ref{psiMap0}), and doing the $%
\delta \theta $ integral, we find the conditions for a solution.

First, we must require 
\begin{equation}
\psi (\theta +\pi )=e^{i\omega _0}\psi (\theta )  \label{PerCond}
\end{equation}
In this case, $\omega _0$ is $0$ or $\pi $, since $\psi $ must be $2\pi $
periodic. Insisting that the leading small terms vanish, we find that $\psi $
must also satisfy the equation 
\begin{equation}
-\frac 1k\psi ^{\prime \prime }+k\epsilon V(\theta )\psi =k\epsilon E_m\psi
\label{psiWKB}
\end{equation}
This is a familiar equation, similar to the equation of a quantum pendulum,
for motion in a periodic potential of strength $k^2\epsilon =\left( k\sqrt{%
\epsilon }\right) ^2,$ and unit Planck. Alternatively, we may take a
potential whose scale is unity and think of Planck's constant as $\hbar =1/(k%
\sqrt{\epsilon }).$ Thus $b=\sqrt{\epsilon }$ naturally appears. This is
such a well known and thoroughly analyzed equation, that the problem can be
considered solved.

If $kb$ is sufficiently small, we can make an ordinary quantum perturbation
expansion about the starting state of zero angular momentum $\psi =$
constant. If $kb$ is larger, some other method must be used.

If $kb$ is large, standard WKB theory makes the Ansatz $\psi (\theta )=\exp
\left( ikbf(\theta )+ig(\theta )+O(1/kb)\right) $ where $df/d\theta =$ $%
f^{\prime }$ $\sim 1$ and $g\sim 1.$ Using this approach gives 
\begin{equation}
f^{\prime }(\theta )=\pm \sqrt{E_m-V(\theta )}  \label{fPrime}
\end{equation}
which is $\pi $ periodic. If $E_m>V$ for all $\theta $, we may quantize $E_m$
by the condition 
\begin{eqnarray}
kb(f(\theta +\pi )-f(\theta )) &=&\pm kb\int_0^\pi d\theta \sqrt{%
E_m-V(\theta )}  \nonumber  \label{RotQC} \\
&=&\omega _o+2\pi m.  \label{RotQC}
\end{eqnarray}
We call this the `rotational' case by analogy with the pendulum.

The usual next order analysis gives $g(\theta )=i\frac 12\ln f^{\prime
}(\theta )$, which is customarily written as a prefactor $\left(
E_m-V(\theta )\right) ^{-1/4}.$

If $E_m<\max V(\theta )$, there will be potential wells, at least two in
this case, with corresponding librational motion. This may be treated at
various levels of approximation. If tunnelling between the wells is
neglected, there will be a quantization condition 
\begin{equation}
kb\int_{\theta _{m-}}^{\theta _{m+}}d\theta \sqrt{E_m-V(\theta )}=\pi (m+\nu
).  \label{LibQC}
\end{equation}
Here, the limits are the angles where the square root vanishes, the
classical turning points, where the leading WKB approximation breaks down.

The Maslov index $\nu $, usually $\frac 12,$ is included. It can be found,
for example, by the usual device of approximating $V$ by a linear function
in the neighborhood of the turning points, finding the Airy function
solution, and using it to interpolate between the WKB solutions away from
the turning point.

The wavefunction, sufficiently inside the turning points, is approximately
given by 
\begin{equation}
\psi =\left( E_m-V\right) ^{-1/4}\sin \left( kb\int_{\theta _{m-}}^\theta d%
\tilde \theta \sqrt{E_m-V(\tilde \theta )}+\frac \pi 4\right)
\label{psi1WKB}
\end{equation}

This gives a two-fold degeneracy, since there are librational levels in two
identical wells. If desired, the exponentially small splitting of these
levels can be estimated.

The condition $E_m=\max V(\theta )$ gives the separatrix of the motion
between librational and rotational motion. Again, the simplest WKB
approximation in the neighborhood of the separatrix can be corrected by well
known if tedious methods.

Is it also straightforward to take into account symmetries which may exist,
for example under $\theta \longleftrightarrow -\theta ,$ or time reversal,
which guarantees that eigenfunctions of Eq.(\ref{psi}) can be taken real.

\subsection{Small angular momentum, $k\epsilon \geq 1,$ $k\epsilon ^{3/2}<<1$%
}

If $k\epsilon \geq 1$ it is not possible to expand the exponential
representing the perturbation in the $T$ operator. We can show, however,
that the WKB solution of the previous section remains valid.

Expanding the phase of the $T\,$ operator about the point $\theta ^{\prime
}=\theta +\pi ,$ as before, we obtain 
\begin{equation}
S(\theta ,\theta ^{\prime })=kL\approx 2k-%
{\textstyle {1 \over 4}}
k\delta \theta ^2+k\epsilon (\Delta R(\theta )+\Delta R(\theta ^{\prime }))
\label{kLa}
\end{equation}
[In Eq.(\ref{kLa}) we replaced $L_0$ by $2$, its stationary value, when
multiplied by $\epsilon .$]

The WKB Ansatz is $\psi =\exp ikbf(\theta ).$ The same prefactor will also
be found, but we ignore it for now.

Returning to Eq.(\ref{psiMap}), and using this Ansatz, we expand all
functions of $\theta ^{\prime }$ about $\theta +\pi .$ I.e. $f(\theta
^{\prime })\approx f(\theta +\pi )+\delta \theta \,f^{\prime }(\theta +\pi )$
to order $\delta \theta $, [since $kb<<k$] and $\Delta R(\theta ^{\prime
})\approx \Delta R(\theta +\pi ),$ since $kb>>k\epsilon $. Doing the
integral reduces Eq.(\ref{psiMap}) to 
\begin{eqnarray}
&&\ \exp [ikbf(\theta )+i\omega ]  \nonumber  \label{E1} \\
\ &=&i\exp \left[ i\left( 2k+(kbf^{\prime })^2/k+k\epsilon V(\theta
)+kbf(\theta +\pi )\right) \right]  \label{E1}
\end{eqnarray}
where $V(\theta )=\Delta R(\theta )+\Delta R(\theta +\pi )$ as before.

For Eq.(\ref{E1}) to hold, the exponents of order $kb$ must combine to give
a constant $\omega _0$, i.e. $f(\theta +\pi )=f\left( \theta \right) +\omega
_0/kb.$ At order $kb^2$ solution is possible provided $(f^{\prime
})^2+V(\theta )$ is a constant, which again we call $E_m$. Thus again 
\begin{equation}
f(\theta )=\pm \int^\theta d\theta ^{\prime }\sqrt{E_m-V(\theta ^{\prime })}.
\label{f}
\end{equation}
The lower limit can be chosen at our convenience. Notice $V(\theta
)=V(\theta +\pi )\Longrightarrow f(\theta +\pi )=f\left( \theta \right)
+const.$

Everything goes through exactly as in the previous section for the
rotational states where $E_m>\max V.$ For $E_m\leq \max $ $V$ we have
librational (or near separatrix) states where leading order WKB fails in the
usual way.

However, we may proceed by almost the standard technique. Namely, for angles
near the turning point, $k\epsilon (E_m-V)$ is small, even if $k\epsilon $
is not. {\em In this region,} therefore, we can expand as in Eqs.(\ref
{TExpand}), (\ref{psiExpand}). The solution in this region interpolates
between the regions where WKB is good and we obtain the usual results,
including prefactors and Maslov indices.

In Fig. 4 we compare the results of this approximation with exact numerical
determination of the wavefunctions. The exact results show the tunnelling
tails and Airy-function-like interpolation at the turning points, which we
have not bothered to calculate in WKB theory.

\begin{figure}[tbp]
{\hspace*{-0.2cm}\psfig{figure=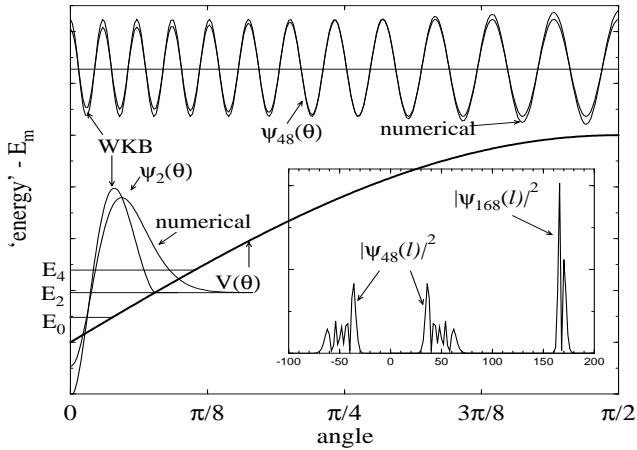,height=8.5cm,width=6cm,angle=270}} {%
\vspace*{.27in}}
\caption[ty]{`Stadium' potential $\left| \sin \theta \right| $ vs angle.
States and potential are symmetric about zero angle. Bound and continuum,
WKB and exact states are shown, with zeroes at WKB `energy' parameter $E_m
\cdot $ $k$$\protect\sqrt{ \epsilon }$ $=42.3$ is fixed. Inset: Angular
momentum representation of continuum state $m=48$ and exact state near
angular momentum $m=168.$}
\label{fig:4}
\end{figure}

We remark that in this case, the results are independent of the size of $%
k\epsilon ,$ that is, there are no corrections to the result of order $%
k\epsilon ,$ but only of order $k\epsilon ^{3/2}.$ Thus we show two
wavefunctions that cannot be distinguished, with the same value of $kb$, but
quite different values of $kb^2,$ one smaller than unity, the other larger.

We remark that the $\delta \theta $ integral is effectively over a width of
order $1/\sqrt{k}.$ However, the center of the effective window of
integration is not at $\delta \theta =0,$ if we take into account the
existence of $f$, but rather at $\delta \theta \simeq 2\sqrt{\epsilon }%
f^{\prime }.$ This shift will be small compared with the width if $\sqrt{%
\epsilon }<<\sqrt{1/k}$, or $k\epsilon <<1.$ Thus the term in $\psi ^{\prime
}$ of Eq.(\ref{psiExpand}) does not contribute for $k\epsilon <<1.$ In the
present section, that condition does not hold, but in the turning point
region where we are expanding, the shift is small because $f^{\prime }$ is
small.

Thus, we conclude that, even if $k\epsilon \geq 1$, the solution is given by
the solution of Eq.(\ref{psiWKB}), using the usual WKB methods.

\subsection{Classical interpretation}

We now give the interpretation of these results in classical terms.

The action $S(\theta ,\theta ^{\prime })$, the phase of the $T$ operator,
generates the surface of section map ${\cal T}(l^{\prime },\theta ^{\prime
})\rightarrow (l,\theta )$ by the equations $l=\partial S/\partial \theta ,$ 
$l^{\prime }=-\partial S/\partial \theta ^{\prime }$. For the distorted
circle billiard, this is explicitly $l=-k(\theta -\theta ^{\prime }+\pi
)/2+k\epsilon \Delta R^{\prime }(\theta );$ $l^{\prime }=-k(\theta -\theta
^{\prime }+\pi )/2-k\epsilon \Delta R^{\prime }(\theta ^{\prime }).$ It is
customary to desymmetrize this map a bit by putting $l=\tilde l+k\Delta
R^{\prime }(\theta ).$ Then the map is 
\begin{eqnarray}
\tilde l &=&\tilde l^{\prime }+k\epsilon V^{\prime }\left( \theta ^{\prime
}\right)  \nonumber \\
\theta &=&\theta ^{\prime }-\pi -2\tilde l/k  \label{circmap}
\end{eqnarray}
This map has been obtained by other methods\cite{BCL,FS}.

If qualitative understanding is the main motivation, $S\,$ may be simplified
while keeping the important physics intact. For example, we may regard Eq.(%
\ref{kLa}) as exactly defining a map which we wish to study classically
and/or quantally. A slight but natural simplification is to use $\delta
\theta =\theta ^{\prime }-\theta .$ It seems natural to take $\Delta R$ as
the simplest possible. Taking $\Delta R=\cos \theta $ as the simplest
possible periodic perturbation yields the well known Chirikov-Taylor or
standard map\cite{Borgo,Chirikov}. Although this choice of $\Delta R$ does
capture most of the phenomena of interest, we find that it is not typical in
certain senses. We return to this point below.

Let the invariant loop be given by a formula $l_{INV}(\theta ).$ This
function will have to be two valued if it is defined for a finite range of
angle, and it can be single valued if it is defined over the full range of
angle. It must satisfy 
\[
{\cal T}(l_{INV}(\theta ^{\prime }),\theta ^{\prime })\rightarrow
(l_{INV}(\theta ),\theta ). 
\]
In terms of the generating function, it must satisfy $l_{INV}(\theta
)=\partial S(\theta ,\theta ^{\prime })/\partial \theta ,$ $l_{INV}(\theta
^{\prime })=-\partial S(\theta ,\theta ^{\prime })/\partial \theta ^{\prime
}.$ Integrating this relation, and calling $F(\theta )=\int l_{INV}(\tilde 
\theta )d\tilde \theta $, we see that 
\begin{equation}
F(\theta )=S(\theta ,\theta ^{*})+F\left( \theta ^{*}\right) +const
\label{CInvMap}
\end{equation}
where $\theta ^{*}(\theta )$ is the angle such that $\partial S(\theta
,\theta ^{*})/\partial \theta ^{*}+F^{\prime }(\theta ^{*})=0.$

If $S(\theta ,\theta ^{\prime })+F\left( \theta ^{\prime }\right) $ appears
as a phase in an integral over $\theta ^{\prime }$, the stationary phase
point is of course $\theta ^{\prime }=\theta ^{*}.$ Thus we see that we find
an approximate invariant loop, $l_{INV}(\theta )\approx k\sqrt{\epsilon }%
f^{\prime }(\theta )=\pm k\sqrt{\epsilon }\sqrt{E_m-V\left( \theta \right) }%
. $ More generally, this is the first term in a series expansion for $%
l_{INV} $. Therefore, $E_m$ is an approximate constant of the classical
motion, which, like the energy, will be quantized.

The loop $l_{INV}$ we have found so far is not invariant in a more accurate
assessment. Rather, it will be mapped into a new loop ${\cal T}%
(l_{INV})\rightarrow l_1(\theta )\neq l_{INV}(\theta ).$ The area enclosed
by the two loops is the same, since ${\cal T}$ is area preserving. It is
possible to estimate the phase space area of points inside $l_1$ but outside 
$l_{INV}.$ In the worst case, the area occupied by such `turnstile' points
is proportional to $\epsilon ^{3/2}$, which is a factor $\epsilon $ smaller
than the area of $l_{INV}$ itself. If the turnstile area is much smaller
than $\hbar $, i.e. if $kb^3<<1,$ it is reasonable to believe that $l_{INV}$
is a good approximation for the purposes of quantum mechanics. Of course, it
may be possible to make a correction to $l_{INV},$ i. e. to find a higher
order approximation to the invariant loop such that the turnstile area is
even smaller. This would then allow us to take $kb^3\geq 1$ and still obtain
good results.

We thus see that the libration states with $E_m<\max V$, are to be
identified with the resonance islands about the stable fixed points. The
rotational states with $E_m>\max V$ are distortions of the unperturbed
invariant loops $l_{INV}=const.$ If $E_m>>\max V,$ this distortion is small.
However, this approximation breaks down at some stage because there will be
higher resonances which must be taken into account.

\subsection{Energy and quasienergy quantization}

We have seen that there is a WKB quantization of the approximate classical
constant of the motion $E_m.$ This leads to an expression for the
quasienergy $\omega $, of Eq.(\ref{psiMap0}) 
\begin{equation}
\omega =\omega (k)=2k+k\epsilon E_m+\omega _0+\pi /2  \label{eigenph}
\end{equation}
Table 1 gives a comparison of numerical quasienergies in comparison with
this formula for a number of states. The method used is that introduced\cite
{FGP} to solve the standard map. Starting with an approximate wavefunction,
a long time series is obtained by applying the $T$ operator repeatedly.
Fourier transform of this series yields the eigenphases and eigenfunctions.
The $T$ operator can be applied very efficiently if it can be factored into
a part dependent only on $\theta -\theta ^{\prime }$, and a part whose phase
is a sum of a function of $\not \theta $ and a function of $\theta ^{\prime
}.$ The latter is not strictly true for the nearly circular billiard, but it
is true to order $b^3.$ Note also that we are comparing our method to the
numerical solution Eq.(\ref{psiMap}), and not to the numerical solution of
the Helmholtz equation. The question of how well the $T$ operator
corresponds to the billiard has already been addressed\cite{TLit}.

The results are quite good, that is, the errors are small compared with the
separation of the levels. The separation is of order 
\begin{equation}
k\epsilon (E_{m+1}-E_m)\simeq \sqrt{\epsilon }  \label{QESep}
\end{equation}

The {\em energy} levels are given by choosing values of $k$ which solve 
\begin{equation}
\omega (k)=2\pi n.  \label{eigenE}
\end{equation}
This has solutions $k=k_{n,m}$ as for the perfect circle. [For $\Delta R=0$, 
$\omega =2\pi n$ reduces to $2k+m^2/k+\pi m=(n-\frac 14)2\pi $ where $m$ is
the angular momentum. This is equivalent to Debye's approximation to
Bessel's function, valid for $k$ large and $m/k$ small.]

For fixed $m$, the variation of $\omega $ with $k$ from the first term, $2k,$
of Eq.(\ref{eigenph}) dominates. Thus we have 
\begin{equation}
k_{n+1,m}\simeq k_{n,m}+\pi .  \label{delK}
\end{equation}

The error in the determination of a given energy level is not, however,
small in comparison with the mean spacing in energy of {\em all} the levels.
The levels so far found are a small fraction of the levels in a given energy
range. There are levels belonging to larger angular momenta and smaller
radial wavenumber is the same range. In terms of $k,$ the level spacing of
all the levels is about $2/k$, which is the order of the size of the errors
committed and is the order of the absolute error of a given level, as is
shown in table2.

\vbox{
\begin{table}[b]
\vspace{0.5cm}
\centering
\begin{tabular}{cccc}
$m$ & $E_m+4/\pi $ & $\omega $ numerical & $\omega $ theoretical \\ 
\hline 
$1$ & $0.2083$ & $5.6199$ & $5.6310$ \\ 
$3$ & $0.4321$ & $5.7701$ & $5.7680$ \\ 
$5$ & $0.6053$ & $5.8726$ & $5.8741$ \\ 
$7$ & $0.7544$ & $5.9661$ & $5.9654$ \\ 
$9$ & $0.8878$ & $6.0465$ & $6.0471$ \\ 
$11$ & $1.0097$ & $6.1222$ & $6.1218$ \\ 
$13$ & $1.1223$ & $6.1904$ & $6.1908$ \\ 
$15$ & $1.1272$ & $6.2553$ & $6.2550$ \\ 
$17$ & $1.3254$ & $0.0317$ & $0.0319$ \\ 
$19$ & $1.4175$ & $0.0885$ & $0.0884$ \\ 
$21$ & $1.5040$ & $0.1411$ & $0.1413$ \\ 
$23$ & $1.5852$ & $0.1912$ & $0.1911$ \\ 
$25$ & $1.6614$ & $0.2376$ & $0.2378$ \\ 
$27$ & $1.7327$ & $0.2814$ & $0.2814$ \\ 
$29$ & $1.7989$ & $0.3217$ & $0.3220$ \\ 
$31$ & $1.8599$ & $0.3592$ & $0.3593$ \\ 
$33$ & $1.9152$ & $0.3924$ & $0.3932$ \\ 
$35$ & $1.9638$ & $0.4243$ & $0.4229$ \\ 
$37$ & $2.0010$ & $0.4461$ & $0.4458$\end{tabular}
\vspace{0.5cm}
\caption{Numerical and theoretical quasienergies, compared for states with
different $m$, but all belonging to the same low angular momentum resonance. 
$\Delta R$ corresponds to the stadium billiard. WKB `energy' parameter is
also given, $k=1000$ and $\epsilon =6.1\times 10^{-4}$ are fixed.}
\end{table}
}

\vbox{
\begin{table}[b]
\vspace{0.5cm}
\centering
\begin{tabular}{cccc}
$n$ & $E_m+4/\pi $ & $k$ numerical & $k$ theoretical \\ 
\hline 
$319$ & $1.0108$ & $998.3207$ & $998.3213$ \\ 
$318$ & $1.0128$ & $995.1784$ & $995.1789$ \\ 
$317$ & $1.0148$ & $992.0358$ & $992.0364$ \\ 
$316$ & $1.0169$ & $988.8934$ & $988.8940$ \\ 
$315$ & $1.0190$ & $985.7509$ & $985.7515$ \\ 
$314$ & $1.0210$ & $982.6085$ & $982.6090$ \\ 
$313$ & $1.0231$ & $979.4660$ & $979.4666$ \\ 
$312$ & $1.0252$ & $976.3235$ & $976.3241$ \\ 
$311$ & $1.0273$ & $973.1810$ & $973.1816$ \\ 
$310$ & $1.0294$ & $970.0387$ & $970.0392$\end{tabular}
\vspace{0.5cm}
\caption{Energies $k$ with different quantum numbers $n,$ but same $m=11$,
computed numerically and found solving Eqs. (\ref{eigenph}) and (\ref{eigenE}
). $\epsilon$ as in Table 1.}
\end{table}
}

It would be nice to have a theory giving the energy levels with absolute
accuracy less than the mean level spacing, of course. Primack and Smilansky 
\cite{SmAcc} have discussed how this might be possible within the framework
of quasiclassics, even though the errors made in arriving at the $T$
operator and doing stationary phase integrals are of order $\hbar .$ This
however involves finding the mean density of levels to better than leading
order accuracy by a separate calculation, and combining it in a particular
way with sums over periodic orbits.

Nevertheless, the results we have obtained are very useful and contain
nearly all that is desired. The energy levels are classified into groups.
Matrix elements of smooth operators are large only between levels in the
same group. The error is small compared with the spacing of the levels in
the same group. The overall statistics of the levels, when found on the
scale of the mean level spacing, is also given correctly. Long range
correlations of energy levels are also given correctly\cite{Zait1}.

\subsection{Wavefunctions and localization}

The `wavefunctions' are also given to good approximation, as we showed above.

Of course, there could be an accidental degeneracy between energy levels
coming from small angular momenta, such as we have discussed above, and
energy levels carrying large angular momenta, which we will calculate below.
Then, the true eigenstates will be some appreciable admixture of large and
small angular momenta. Lazutkin's rigorous results also allow for this
possibility\cite{Lazutkin}. However, the matrix elements of the $T$ operator
between widely different angular momentum states is exponentially small.
This means that such accidental degeneracies will be rare.

In the presence of symmetries, there will be degeneracies which are exact in
the absence of the exponentially small coupling. An example is the states
associated with the periodic wells discussed above. In the presence of time
reversal invariance, the states associated with positive angular momenta are
degenerate with those of the corresponding negative angular momenta, again
with very small splitting due to tunnelling between the two momentum
regions. If desired, this splitting can be estimated within the framework of
our theory. This is not entirely trivial as it requires a study\cite{Zait1}
of `resonance assisted tunnelling' which is an analog of `chaos assisted
tunnelling'\cite{CATun}.

The wave function so far found contains only small angular momentum
components. This can be seen explicitly in angular momentum representation,
determined schematically from the integral 
\begin{equation}
\psi _l=\int d\theta \left( E_m-V(\theta )\right) ^{-\frac 14}\exp \left( ik%
\sqrt{\epsilon }f(\theta )-il\theta \right) .  \label{AngMomRep1}
\end{equation}
If $k\sqrt{\epsilon }>>1,$ stationary phase may be employed. Then, it is
found that only the neighborhood of the angles solving 
\begin{equation}
k\sqrt{\epsilon }f^{\prime }(\theta )=k\sqrt{\epsilon }\sqrt{E_m-V(\theta )}%
=l  \label{AMCond}
\end{equation}
contribute appreciably. A range of $l$ centered roughly at $k\sqrt{\epsilon
E_m}$ and of width of order $k\sqrt{\epsilon \max V}$ can satisfy this
condition, so $k\sqrt{\epsilon \max V}$ is a sort of localization length for
this kind of wave function. Note that only librational states of the lowest
resonance have much overlap with zero angular momentum.

Outside this range, the angular momentum components of the wave function
decay exponentially for smooth $V$. For the stadium case, $V=\left| \sin
\theta \right| $, it may easily been seen that the decay is as $l^{-4}$,
rather than exponential. This makes it necessary to use nonstandard
definitions of the localization length\cite{BCL,Borgo,CP} if statistical
results are to be calculated.

Carrying out the integral of Eq.(\ref{AngMomRep1}) gives an expression 
\begin{equation}
\psi _l=\sum_{\theta _a}\left| V^{\prime }(\theta _a)\right| ^{-\frac 12%
}\exp i\left( k\sqrt{\epsilon }f(\theta _a)-l\theta _a+\nu _a\frac \pi 2%
\right)  \label{AngMomRep2}
\end{equation}
where $\theta _a$ are the solutions of Eq.(\ref{AMCond}).

Notice that, if as expected, $V^{\prime }$ has different signs at the
different $\theta _a,$ there will be an additional Maslov phase index $\nu
_a $ distinguishing the two solutions. Also notice that the singularity of
the prefactor at the classical turning point in angle, where $V\left( \theta
\right) =E_m$ has disappeared. Instead, the angular momentum regions which
make $V^{\prime }$ vanish, which are the turning points in the angular
momentum representation, are singular, and should be treated by a technique
going beyond the first WKB approximation. This shift of the region of
breakdown of the leading WKB approximation under Fourier transform is the
basis of Maslov's treatment\cite{Maslov} of this subject.

The solutions thus found are {\em localized} in angular momentum space. From
the point of view of KAM theory and perturbation theory, this is
unremarkable. First, the KAM theory predicts that the classical motion will
be confined to much the same region. Second, matrix elements of the $T$
operator in the angular momentum representation are small away from the
diagonal. In fact, consider $T_{ll^{\prime }}=(2\pi )^{-1}\int \int d\theta
d\theta ^{\prime }e^{-il\theta }T(\theta ,\theta ^{\prime })e^{il^{\prime
}\theta ^{\prime }}$ where $(l-l^{\prime })/k$ is of order unity. The $%
\theta ^{\prime }$ integral will be stationary near $dS_0(\theta -\theta
^{\prime })/d\theta =l^{\prime },$ and the $\theta $ integral will be
stationary near $dS_0(\theta -\theta ^{\prime })/d\theta =l.$ Restricting
consideration to the standard case where there is only one stationary point,
we see that the integral will be exponentially small.

Such a band diagonal matrix $T_{ll^{\prime }}$ is studied in localization
theory\cite{FGP} and is effectively the subject of the papers\cite
{BCL,FS,Borgo,CP} cited earlier. Of course, since the long time behavior is
diffusive, in the non KAM cases, it was not obvious exactly the mechanism
for localization. If stationary phase arguments are sufficient to get the
width of the band, as we have assumed, the problem is solved. The cited
papers are concerned with the non KAM case that $k\epsilon ^M$ is
sufficiently large that stationary phase breaks down.

It should be emphasized that our whole theory is based on the assumption,
validated self consistently, that the eigenstates are localized in angular
momentum.

Although it is usually not worth the effort, the full two dimensional
wavefunction can also be found in WKB approximation.

\subsection{Large angular momenta}

We turn to a discussion of the problem of larger angular momenta. As we have
just remarked, a large angular momentum state cannot communicate directly
via the $T$ operator with its negative angular momentum counterpart. It's
true that time reversal symmetric systems will be symmetric under $%
l\rightarrow -l$, so that true eigenstates will be a mixture of these
states. However, the even and odd eigenstates will be practically
degenerate, and we can consider one sign of angular momentum only to get the
main effect.

Suppose we wish to find eigenstates centered near angular momentum $l.$ One
way is to reduce the problem to the one just solved by making a sort of
gauge transformation, in which the eigenstates $\psi \rightarrow e^{il\theta
}\tilde \psi $, $T\rightarrow e^{-il(\theta -\theta ^{\prime })}\tilde T,$
and $\tilde T\tilde \psi =\tilde \psi .$ We may then assume that $\tilde \psi
$ is slowly varying.

There are some complications in this case as compared with the case where $%
l=0.$ First, to leading order, the stationary phase point is given by $%
dS_0(\theta -\theta ^{\prime })/d\theta =l$, namely at $\theta -\theta
^{\prime }=-\Theta _l,$ say. However, $\Theta _l$ has no particular
relationship to $2\pi $, that is, it will be an irrational multiple of $2\pi
,$ unless $l$ is specially chosen. (The negative sign in the definition of $%
\Theta _l$ is purely for orthographic convenience.) If $l\,$ is an integer,
this cannot happen except at special symmetry values such as $l=0.$

There are two possibilities. Either $\Theta _l$ is very close to a rational
multiple of $2\pi ,$ $\Theta _{pq}=2\pi p/q,$ where $q\,$ is not too big, or
it is sufficiently far from such a multiple. It is the job of the theory to
tell us what `very close', `not too big' and `sufficiently far' mean in
detail. The resonant case is the one where $\Theta _l$ is `close' to a
rational, the KAM-set case has $\Theta _l$ `far' from a rational.

\subsubsection{Resonant case}

The first case we deal with is that of a resonance of period $q$, which
involves destruction of the original invariant tori and their replacement by
new tori with a different topology. These are represented on the surface of
section by a period $q$ island chain. The second, nonresonant, case deals
with invariant tori which are not much distorted by the perturbation.

In this case we consider states whose angular momentum components are not
too far from $l_{pq}$ where the leading stationary phase point corresponding
to $l_{pq}$ is $\Theta _{pq}.$ Note that $l_{pq}$ is not generally integer,
so that $\tilde \psi $ will not be $2\pi $ periodic.

The Ansatz made before, 
\begin{equation}
\tilde \psi =\exp (ikb\tilde f(\theta ))  \label{psitwid0}
\end{equation}
does not lead to a solution. Indeed, expanding $\tilde T,\,\tilde f$ about $%
\Theta _l,$ we do the $\theta ^{\prime }$ integral and find 
\begin{eqnarray}
&&\ \exp \left[ ik\epsilon \left( c\tilde f^{\prime }(\theta +\Theta
_{pq})^2+d(\Delta R(\theta )+\Delta R(\theta +\Theta _{pq}))\right) \right] 
\nonumber  \label{Psimap} \\
\ &=&e^{-i\alpha }e^{i\omega }\exp (ikb(\tilde f(\theta )-\tilde f(\theta
+\Theta _{pq})).  \label{psiMap}
\end{eqnarray}
Here $c$ is determined from the second derivative of $S_0.$ The coefficient $%
d$ for the billiard is $d=c^{-2}=\sqrt{1-l^2/k^2}.$ $\alpha $ is the phase
obtained by doing the $\theta ^{\prime }$ integral. For the case defined by
Eq.(\ref{LT}), $\alpha =\frac 12\pi +2\sqrt{k^2-l^2}-2l\cos ^{-1}(\frac lk).$
In this case $\Theta _{pq}=-2\cos ^{-1}(l/k)$, for positive $l.$ To satisfy
Eq.(\ref{psiMap}) we must require 
\begin{equation}
\tilde f(\theta +\Theta _{pq})=\tilde f(\theta )+const  \label{fCond0}
\end{equation}
and 
\begin{equation}
c^2(\tilde f^{\prime })^2+d(\Delta R(\theta )+\Delta R(\theta +\Theta
_{pq}))=const.  \label{fpC0}
\end{equation}
These two conditions cannot in general be simultaneously fulfilled. Eq.(\ref
{fCond0}) implies that $\tilde f^{\prime }$ is $q$-periodic, i.e. periodic
with period $\Theta _{pq},$ but only for special choice of $\Delta R$ will
the terms in $\Delta R$ in Eq.(\ref{fpC0}) be $q$-periodic, if $q\neq 2.$

We therefore improve the Ansatz and take 
\begin{equation}
\tilde \psi =\exp [ikd(bf_1(\theta )+b^2f_2(\theta )+\ldots ).
\label{psitwid}
\end{equation}
Introducing $d$, characterizing $S_0^{\prime \prime }(\Theta _{pq})$,
simplifies later formulas. Treating $b^2f_2$ as slowing varying, in this
order of the calculation, we find the conditions 
\begin{equation}
f_1(\theta +\Theta _{pq})=f_1(\theta )+const  \label{fCond}
\end{equation}
and 
\begin{equation}
f_2(\theta )-f_2(\theta +\Theta _{pq})=f_1^{\prime 2}+\Delta R(\theta
)+\Delta R(\theta +\Theta _{pq})-E_m  \label{fDiff}
\end{equation}
where $E_m$ is a constant to be determined.

Let us define the q-average of a function of angle by $\bar F_q(\theta )=%
\frac 1q\sum_{r=1}^qF(\theta +r\Theta _{pq}).$ The Fourier components of $%
\bar F_q$ are just those of $F$ with all components whose index is not
divisible by $q$ discarded. Making a q-average of Eq.(\ref{fDiff}) leads to 
\begin{equation}
f_1^{\prime }=\pm \sqrt{E_m-\bar V_q(\theta )}  \label{f1pb}
\end{equation}
where $\bar V_q=2\overline{\Delta R}_q.$

The function $f_2$ cannot be completely determined at this level, since an
arbitrary q-periodic function can be added to it and not affect Eq.(\ref
{fDiff}). Calling $V_{pq}=\Delta R(\theta )+\Delta R(\theta +\Theta _{pq})-%
\bar V_q$ we can solve for the remaining part of $f_2.$ This function has no
Fourier components whose index is divisible by $q.$ The solution of $%
f_2(\theta )-f_2(\theta +\Theta _{pq})=V_{pq}$ can be made by Fourier
transform of the equation. An explicit solution can also be written 
\begin{equation}
f_2(\theta )=\frac{-1}q\sum_{r=0}^{q-1}rV_{pq}(\theta +r\Theta _{pq}).
\label{f2}
\end{equation}

This suggests making the following transformation. Let $\tilde \psi
=e^{ik\epsilon df_2}\hat \psi $ and consider the corresponding operator $%
\hat T$ which may be approximated by 
\begin{eqnarray}
\hat T(\theta ,\theta ^{\prime }) &\simeq &\sqrt{\frac{kd}{2\pi i}}\exp
(2ikd-\frac{ikd}4\delta \theta ^2)  \nonumber  \label{THat} \\
&&\times \exp (ikd\epsilon (\Delta R(\theta )+\Delta R(\theta +\Theta _{pq})
\nonumber \\
&&+f_2(\theta +\Theta _{pq})-f_2(\theta )))  \label{THat}
\end{eqnarray}
The second exponential is $\exp (ikd\epsilon \bar V_q).$ If $kd\epsilon $ is
small, we expand $\hat \psi $ as in Eq.(\ref{psiExpand}) above. Thus, $\hat 
\psi $ solves the periodic potential problem with potential $\bar V_q$ and
effective Planck's constant $1/kd\sqrt{\epsilon }.$

\begin{figure}[tbp]
{\hspace*{-0.2cm}\psfig{figure=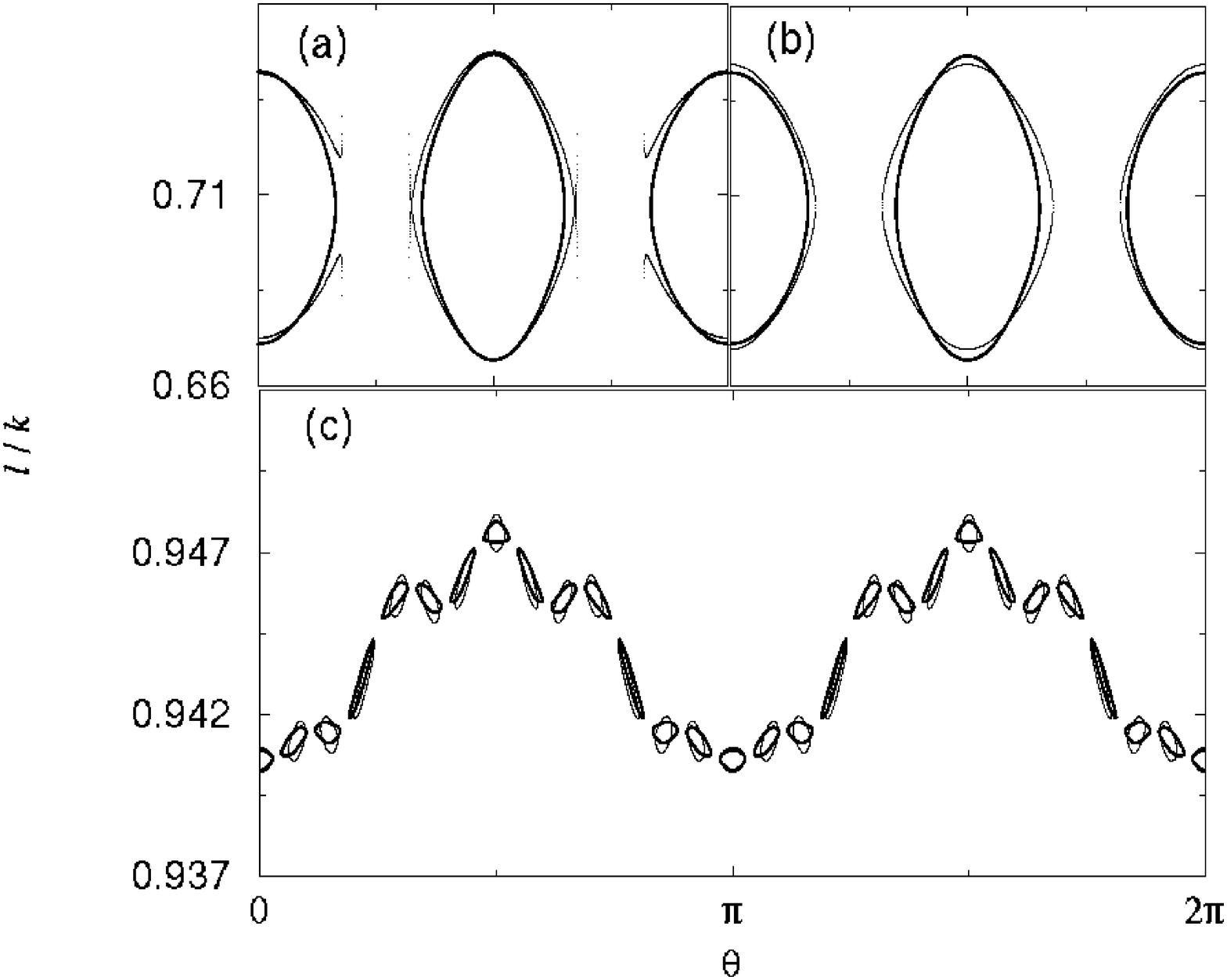,height=6cm,width=8.5cm,angle=0}} {%
\vspace*{.13in}}
\caption[ty]{Approximate, to order $\epsilon$, and exact invariant loops at
the fourth order (b) and twenty-eighth order (c) resonances. The value of $%
\epsilon $ is $0.05$ in (a), (b) and $0.01$ in (c), not very small. $\eta =
0.7$ in (a), (b) and $0.19$ in (c). Thick lines are exact invariant curves,
thin lines are theory. In the fourth order case we have verified that the
failure of the approximation is largely due to the absence of the order $b^3$
contributions - in (a) theoretical curve including $b^3$ order is plotted.
Higher period resonances have relatively, although not absolutely, bigger
corrections from higher orders in $\protect\sqrt{\epsilon }$. }
\label{fig:5}
\end{figure}

If $kd\epsilon $ is of order unity or larger, the argument goes through as
before, so that we have obtained the solution to the resonant case. This
also gives the prefactors, as before.

We show in Figs. 5a,b 5c some invariant loops for higher resonances $pq=1,4$
and $pq=3,28$ respectively. The numerical loops obtained by iterating the SS
map are compared with the theoretical $l_{INV}=l_{pq}+bf_1^{\prime
}+b^2f_2^{\prime }$ .

There are several things of note. First, the contribution of $f_2$ becomes
relatively more important for higher $q.$ This is for two reasons. Most
obviously, $(\epsilon \bar V_q)^{1/2}$ systematically becomes smaller with
increasing $q$, and eventually becomes smaller than $\epsilon ,$ so that the
small resonance islands `float' on a relatively large `wave' supplied by $%
f_2.$ Second, $f_2$ can itself become rather large. This can be seen from
formula (\ref{f2}), which can become fairly big if the terms in the sum add
`in phase'. This effect is most pronounced near a large resonance island,
e.g. at $q=2$. Then the term $\epsilon f_2^{\prime }$ for $p/q$ slightly
greater than $\frac 12$ with $q$ large, must be almost as big as $\sqrt{%
\epsilon }f_1^{\prime }$ for $q=2.$

Another remark is that higher order terms more generally are relatively more
important for large $q$. This can be seen in the results (Fig. 5c). These
deviations are apparent in this case of rather large $b=0.1,$ $b^2=0.01$, $%
b^3=0.001.$ Again this is because $\bar V_q$ is becoming smaller, while the
higher order corrections presumably remain roughly constant. The observed
deviations are apparently of order $b^3.$ Of course, if $kb^3$ is small, the
quantum results won't be affected by the deviations.

\begin{figure}[tbp]
{\hspace*{-0.2cm}\psfig{figure=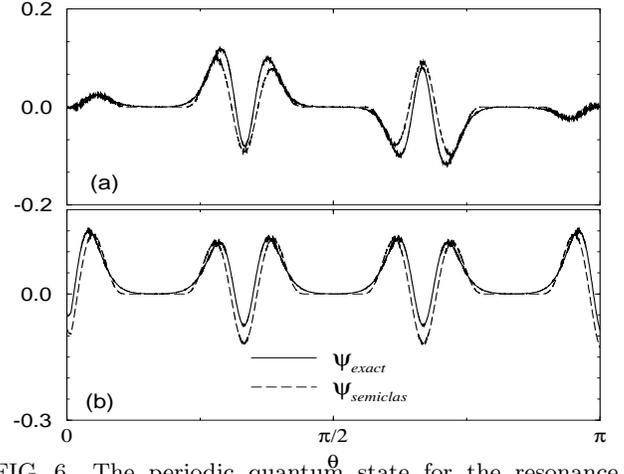,height=8.5cm,width=6cm,angle=270}} {%
\vspace*{.13in}}
\caption[ty]{The periodic quantum state for the resonance $pq=1$,$3$.
Theoretical wavefunction is $\cos \left( kdbf_1\right)$ $\cdot$
$ \sin \left( l\theta
+kdb^2f_2\right) $; in (a), (b) we integrated out exact and semiclassical
functions locally with respectively $\sin \left( l\theta \right) $, $\cos
\left( l\theta \right) $. $k=4032$, $\epsilon =6.79\times 10^{-4}$.}
\label{fig:6}
\end{figure}

The case $pq=3,28$, shows an impressive set of wriggles coming from $%
f_2^{\prime }$ which are quite well predicted theoretically. We have not
tried to find the `best' value $E_m$ corresponding to a given classical
numerical $l_{INV},$ except by trial and error. This may contribute somewhat
to the discrepancies. The islands are of different shape, however, and vary
in height from one island to the next. This effect cannot be reproduced to
the order which we have worked.

In Fig. 6 we also show the quantum states corresponding to the resonance $%
pq=1,3$. What is plotted is $\tilde \psi ,$ that is, the rapidly varying
factor $\exp il\theta $ is removed.

\subsubsection{Quantization conditions}

The quantization conditions are somewhat modified, since the central angular
momentum $l_{pq}$ is not usually integer. We require that $\exp
(il_{pq}\theta +ikdbf_1(\theta ))$ be $2\pi $-periodic. For the rotational
levels, $E_m>\max \bar V_q$ we have a modified quantization condition 
\begin{equation}
\int_0^{2\pi }d\theta \sqrt{E_m-\bar V}=2\pi (m+\delta )/kbd  \label{QC2}
\end{equation}
where $-\delta $ is the fractional part of $l_{pq}.$

If there are solutions of 
\begin{equation}
kbd\int_{\theta _{m-}}^{\theta _{m+}}d\theta \sqrt{E_m-\bar V_q(\theta )}%
=\pi (m+\nu )  \label{LibQC2}
\end{equation}
where the integral is between turning points straddling a minimum of the
potential, $\nu $ again being the Maslov index, there will be librational
solutions repeated $q$ times. There are then $q$ nearly degenerate states,
and the phase shift from one well to the next will be slightly different
than $\Theta _{pq}$ in such a way as to make the total wave function $2\pi $%
-periodic.

These quantization conditions make the area of the loops an integer number
of Planck's plus the correction due to Maslov indices. Thus, it is
equivalent to EKB quantization.

\subsubsection{When the resonant solution is applicable and needed}

The crucial scale is usually provided by $\bar V_q$. We assume without loss
of generality that $\Delta R$ and thus $\bar V_q$ have vanishing mean.

If librational states exist, then we clearly need the resonant solution.
This requires $\bar V_q\geq q^2/k^2\epsilon .$

If $\bar V_q$ is very small, the smallest rotational $E_m$ is of order $%
E_m\sim (kbd)^{-2}.$ Thus, if $\bar V_q\geq (kbd)^{-2}$ we cannot neglect
it. If $\bar V_q$ is much less than this value we can neglect it and the
states in this angular momentum neighborhood are candidates to be treated by
methods valid for the KAM set.

We remark on several cases which are distinguished by the way the Fourier
components of $\Delta R$ behave.

If $\Delta R$ is a pure low harmonic, or sufficiently close to one, e.g. $%
\Delta R=\cos \theta $ or $\Delta R=\cos 2\theta ,$ then $\bar V_q=0$ for $%
q>2.$ Of course, it is immediately recognized that these cases are to first
approximation a shifted circle and an integrable ellipse, respectively, so
that they can be transformed to a perturbation about an integrable system
with perturbation parameter $\epsilon ^2$ rather than $\epsilon .$

From our perspective, if $\bar V_q$ vanishes or is sufficiently unusually
small, one should determine if some higher power of $\epsilon $ gives
resonant behavior. This is not hard to do in the simplest cases, but is
tedious for more complicated cases. For example, the order $q$ resonance of
the standard map for small $\epsilon $ will have width parameter
proportional to $\epsilon ^{q/2},$ since the effective resonant coupling is $%
(\epsilon \cos \theta )^q.$ Cases where the Fourier expansion of $\Delta R$
is a finite Fourier series will have a rather complicated condition which is
difficult to treat in generality.

Analytic perturbations usually have Fourier coefficients which drop off
exponentially with index. This leads to an estimate, $\bar V_q\sim \eta ^q.$
If $\eta >\epsilon ,$ we can expect that the treatment given above is
adequate and the leading order expression for the resonance is proportional
to $\sqrt{\epsilon }.$ If not, we either have to work out the higher order
effects or take $k$ sufficiently small that we can ignore the resonance
entirely.

Finally, there are the cases with nonanalytic $\Delta R$, whose Fourier
coefficients drop off as a power, say. The billiard $\Delta R$ $=\left| \sin
\theta \right| $ has $\bar V_q\sim q^{-2}$, for example, where KAM theory
breaks down completely. We can generally expect our leading order solution
to be valid for these cases. In this case, the entire phase space is best
regarded as filled with resonances\cite{Cantori}, leaving no KAM set.

\subsubsection{Close to a large resonance}

The following issue is to some extent still unresolved. Suppose $\bar V_q$
is sufficiently large that Eq.(\ref{QC2}) holds nontrivially. Consider, for
example, the resonance labelled $pq=7,16$ in Fig. 2. For values of $E_m>\max 
\bar V_q$ the predicted invariant loops are well described by the low order
resonance formula, although they don't give higher order resonances.
Assuming $E_m$ large enough to expand in $\bar V_q$, the invariant loop is
given by $bf_1^{\prime }+b^2f_2\approx b\sqrt{E_m}(1-\frac 12\bar V%
_q/E_m)+b^2f_2$. This may be shown to approximate the nonresonant solution,
also called $f_2$ but for a nonresonant value of $l,$ found in the next
section\cite{Zait1}. However, even if one is quite close to the separatrix,
where one expects the resonant formula to be best, the resonant or
nonresonant formula at the appropriate $l$ value is not bad. Even the
formula for a higher order resonance which seemingly lies in a region
strongly affected by the big resonance does qualitatively quite well,
although there are some small discrepancies.

This seems to present a contradiction. Close to the big resonance, e.g. $%
p=1,q=2,$ a secondary island chain from a high order resonance follows well
a curve given by $b\sqrt{E_m-\bar V_2}$ {\em and also a curve with seeming
smaller variation }$\epsilon f_2$, corresponding to a resonance $pq.$
However, according to Eq.(\ref{f2}), $f_2$ can become quite large for large $%
q$, which can compensate for the fact that $\epsilon <<\sqrt{\epsilon }.$

\subsubsection{Nonresonant case, quantization of the KAM set}

In this case, we start with a value of integer angular momentum $l$, such
that $\Theta _l$ is not resonant. There is no term $f_1$ and the solution
begins with $f_2.$ The condition for $f_2$ is 
\begin{equation}
f_2(\theta +\Theta _l)-f_2(\theta )=\Delta R(\theta +\Theta _l)+\Delta
R(\theta ).  \label{f2N}
\end{equation}
This may be solved in terms of Fourier components of $\Delta R$ , denoted $%
\Delta R$ $_r.$ The result is 
\begin{equation}
f_2(\theta )=\frac 1{2\pi }\sum_r\frac{e^{ir\Theta _l}+1}{e^{ir\Theta _l}-1}%
\Delta R_re^{ir\theta }.  \label{f2F}
\end{equation}

This sum will be well behaved, provided $\Delta R_r,$ drops off sufficiently
rapidly, and is small when the denominator is small. The denominator is
small, when $r=q$ and $\Theta _l$ is close to a rational $\Theta _{pq},$
with $q$ not too large. The denominator then approximates $i(q\Theta _l-2\pi
p).$ If this condition does not hold, the resonance treatment of the
previous sections must be employed.

At this order of the calculation, there is no shift in the energy levels for
the KAM set states.

\begin{figure}[tbp]
{\hspace*{-0.2cm}\psfig{figure=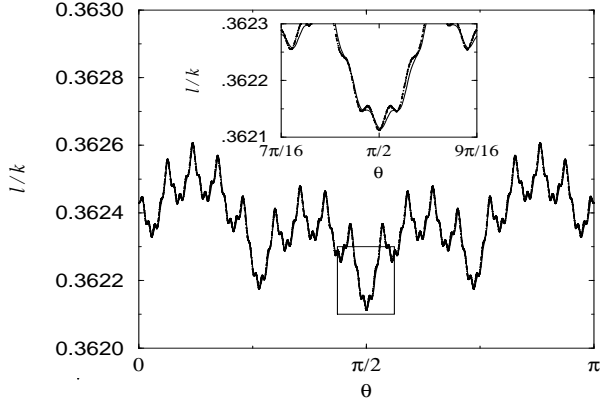,height=8.5cm,width=6cm,angle=270}} {%
\vspace*{.13in}}
\caption[ty]{The golden mean invariant torus, approximated to order $%
\epsilon $ [solid line $\cos \frac {\Theta _l} 2 + \epsilon f_2 (\theta)$],
and numerically exact [classical map represented by dots; 20,000 mappings].
Parameters are $\epsilon = 1 \times 10^{-4},\eta = 2 \times 10^{-2}.$ The
inset enlarges a cusp-like portion of the figure to display the degree of
failure of the approximation. This torus has $\Theta _l=2\pi w=\pi (\protect
\sqrt{5}-1)$, and it is `farthest' from a low order rational. Such a torus
is not really smooth, and is not describable by a convergent power series.
It is `transversally smooth', which is enough to control mathematically its
properties\cite{Lazutkin}. One can pick out the closest resonances, e.g.
5/8, 8/13, etc. given by the Fibonacci series. Presumably, as $\eta $ is
decreased, this is the last torus remaining in the KAM set. Note that the
new tori at the resonances are relatively smooth. Presumably, on the fine
scale of $\epsilon ^{1/\epsilon }$ they would also have such a structure.}
\label{fig:7}
\end{figure}

\begin{figure}[tbp]
{\hspace*{-0.2cm}\psfig{figure=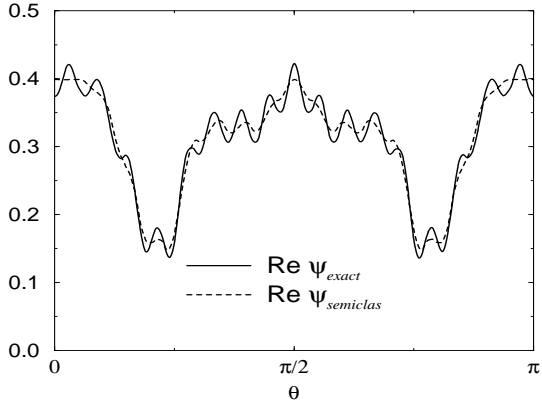,height=8.5cm,width=6cm,angle=270}} {%
\vspace*{.13in}}
\caption[ty]{Exact and approximate wavefunctions for the parameters of Fig.
7. The factor $e^{il\theta }$ has been removed and the real part of the
wavefunction is shown.}
\label{fig:8}
\end{figure}

Formula (\ref{f2F}) has been tested numerically. In Fig. 7 we plot $%
f_2^{\prime }(\theta ),$ the derivative of Eq.(\ref{f2F}) and compare it
with the numerical invariant loop, obtained by propagation of an orbit. It
is apparent from this that the invariant loops are not totally smooth and
featureless. KAM theory controls the singularity of these curves, although
they cannot be represented by a convergent power series. The quantum state
that corresponds to the same KAM\ set was build and compared to the
numerical wavefunction in Fig. 8.

\subsection{Higher order calculations}

We briefly mention how one goes to higher order in the calculation. Some
details are put in the Appendix. We confine attention to the resonant case.

Basically, one makes the Ansatz

\begin{equation}
\psi =\exp \left[ i\left( l_{pq}\theta ^{\prime
}+k(bf_1+b^2f_2+b^3f_3+\ldots )\right) \right]  \label{A2}
\end{equation}
Instead of expanding $S_0$ to second order in $\delta \theta =\theta
^{\prime }-\theta -\Theta _{pq},$ $f_1$ to first order, $f_2$ to zeroth
order, and neglecting $f_3,$ we expand, step by step, $S_0$ to the $M$'th
order, $f_1$ to the $M-1$'th order, ..., $f_M$ to the zeroth order, and
neglect $f_{M+1},$where $M\geq 2,$ and $kb^{M+1}<<1.$ The perturbation,
which we may call $\epsilon S_2(\theta ,\theta ^{\prime })+\epsilon
^2S_4(\theta ,\theta ^{\prime })+...$ is expanded according to the same
rule. The order $r$ calculation determines the non-periodic part of $f_r$
and the periodic part of $f_{r-1}.$

The resulting integral is not done exactly, but by stationary phase in the
neighborhood of $\delta \theta =0.$ The stationary point is regarded as
shifted by $\Delta \Theta _{pq}$ which can be expressed as a power series in 
$b.$ The $b^r$ term in this shift is bigger than the width, $1/k,$ of the
effective region in the $\delta \theta $ integral if $kb^r>>1.$ A sample
calculation is done in the Appendix.

There are interesting questions about how this method breaks down if there
are unusually large derivatives of $\Delta R$, but we shall not address them
in this paper.

\section{Other integrable systems}

There are an infinite number of integrable systems whose perturbations can
be studied. Another interesting billiard is the rectangular billiard. As an
example of a perturbed rectangular billiard, consider the trapezoidal
billiard, whose sides consist of the $x-$axis, $0\leq x<a,$ the vertical
segments, $0\leq y<b-\epsilon ,$ $(x=0),$ $0\leq y<b+\epsilon ,$ $(x=a),$
and the slightly sloping segment connecting the upper ends of the vertical
segments.

The results are much the same as before. We just mention what is found for $%
\epsilon \bar V_q.$ This potential is periodic in $x$, with period $2a/q.$
In one period, the potential is an isosceles triangle `tent', of zero mean,
and peak absolute value $\epsilon /q^2.$ Note that $x$ is continued outside
the physical region.

In fact, for any reflection symmetric perturbation $\Delta R$ of the
circular billiard, there is a corresponding perturbation of the rectangular
billiard, in which only one side of the billiard is perturbed. The leading
orders are almost the same in the two problems, except that the parameter $d$%
, having to do with the second derivative of the integrable action $S_0$ is
unity in the rectangular billiard. Higher orders are different, as well,
which may be helpful in studying those effects.

\section{No small parameter}

In this section we briefly take a case where some states of a system can be
found, even though there is no small parameter. Consider the standard
stadium billiard, which has a straight side of length $2a$, as in Fig. 9.
The side $a$ is arbitrary, but not too small. We want to study the `bouncing
ball' states which have low linear momentum {\em parallel }to these sides,
but a high {\em perpendicular} momentum.

\begin{figure}[tbp]
{\hspace*{0.5cm}\psfig{figure=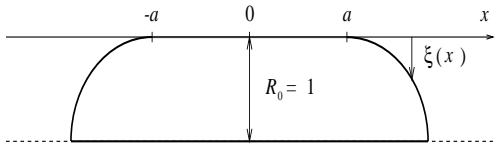,height=6.5cm,width=3.5cm,angle=270}} {%
\vspace*{-.4in}}
\caption[ty]{Half of stadium billiard with long straight side. Notations as
in the section iv.}
\label{fig:9}
\end{figure}

Reduce the billiard to its upper half by symmetry. Take as nominal
integrable system the infinite channel of width $R_0=1$ as shown. We take as
SS the upper boundary of the billiard, and use position $x$, measured from
the symmetry point, on the upper channel to label position on the surface of
section. The billiard boundary is described by distance $\xi $ from the
upper nominal channel, $\xi (x)=0,$ $\left| x\right| <a,$ $\xi (x)\approx
(\left| x\right| -a)^2/2,$ $\left| x\right| >a.$ The integrable action is,
for states odd under reflection about the horizontal symmetry line, 
\begin{equation}
\frac{S_0(x-x^{\prime })}\hbar =2k\sqrt{1+\frac 14(x-x^{\prime })^2}\approx
2k+\frac k4(x-x^{\prime })^2.  \label{S0Stad}
\end{equation}
States even about this line have the same action with an additional term $%
\pi \hbar ,$ that is, the $T$ operator has an additional overall negative
sign for the even states. The change of sign of the second derivative, as
compared with the circle, means that the attractive and repulsive regions of
the potential are reversed in comparison with the previous case. We can
approximate $S_2$ by 
\begin{equation}
S_2(x,x^{\prime })=-k(\xi (x)+\xi (x^{\prime })).  \label{S2Stad}
\end{equation}
Put 
\begin{eqnarray}
V(x) &=&S_2(x,x)=0;\text{ }\left| x\right| <a  \nonumber \\
&=&(\left| x\right| -a)^2;\left| x\right| >a  \label{VStad}
\end{eqnarray}

We look for a solution 
\begin{equation}
\psi =\exp (ikf(x))\text{ }  \label{psiStad}
\end{equation}
where $f$ is slowly varying. One finds, upon approximating $f(x^{\prime
})=f(x)+f^{\prime }(x)(x^{\prime }-x)$ 
\begin{equation}
f^{\prime }(x)=\pm \sqrt{E_m-V(x)}  \label{fpStad}
\end{equation}
Thus, we look for a solution of a particle in the potential well $V(x),$ and
with effective Planck's constant $1/k.$ If $E_m$ is sufficiently small, then 
$f$ varies slowly, and the expansion is justified. The possibility of an
emergent small parameter, such as $E_m$ allows us to use a perturbation
approach.

It's clear that a sufficient number of classical orbits giving a localized
wavefunction is a precondition for special solutions of this kind. This can
happen if the classical orbits are nearly nonisolated. That will happen
systematically near stable periodic orbits.

Some further remarks are made in an Appendix.

\section{Summary}

Weakly perturbed integral Hamiltonian systems in two dimensions have been
much studied. Up to now, the methods employed, although effective
numerically, did not lend themselves to visualization and simple formulas.
By using surface of section techniques as generalized to quantum systems by
Bogomolny, we obtain results very much like familiar results from WKB
theory, at least in the leading orders.

The relation of the perturbation parameter $\epsilon $, to the dimensionless
wavenumber $k,$ or equivalently to $1/\hbar $ is important. The size of $%
k\epsilon $, the number of wavelengths in the change of a billiard boundary,
is not especially important, for resonant states. Rather, the main parameter
is $k\sqrt{\epsilon }$ , which determines the Shuryak border demarcating the
parameter range where ordinary quantum perturbation theory applies and where
it fails. Results are quite simple if $k\epsilon ^{3/2\text{ }}$ is small.
In that case, the derivatives of the perturbation do not enter directly into
the solution. For nonresonant states, the size of $k\epsilon $ determines
whether ordinary perturbation is valid or not.

The results can also be regarded as giving expressions for the invariant
loops, the intersections of invariant tori with the surface of section.
Knowledge of these, rather than of the whole invariant torus structure is
sufficient to quantize the system, quasiclassically.

\section{Future work}

Several avenues for future work are underway. It would be interesting to see
how the perturbation approach breaks down in some detail, in cases such as
the stadium billiard. Working out and verifying higher order effects in
detail will be of some interest. It would be well to extend the method to
arbitrary surfaces of section rather than the special ones we have used. A
number of interesting problems are raised by cases such as $\Delta R=\cos
2\theta .$ Other perturbation systems would be interesting to study, for
example, coupled quartic oscillators\cite{Ullmo}. It would be well to extend
the method to weakly coupled harmonic oscillators. A general theory of
whispering gallery orbits in sufficiently smooth convex billiards\cite
{Tanner,ABGOP}, might be possible, but would require a better $T$ operator
than we have used above, since diffraction effects become important for
short classical skips between nearby boundary points. Extension to $d>2$ may
be in principal possible but in practice is likely to be difficult. There
are interesting numerical results, for example in ray splitting billiards%
\cite{ABGOP} that suggest it might be possible to solve the case that {\em %
two} or more orbits contribute to a specific $T$ operator, requiring a sort
of matrix version of our theory. Finally, it could be hoped to use this
technique to study some or even most states in pseudointegrable systems\cite
{pseudo}, which have nonisolated orbits which however do not lie on a simple
torus in phase space. Many of these projects are underway\cite{Zait1}.

\section{Acknowledgments}

Supported in part by NSF DMR 9624559 and the U.S.-Israel BSF 95-00067-2. We
thank the Newton Institute for support and hospitality in the early stages
of this work. Many valuable discussions with the organizers and participants
of the Workshop `Quantum Chaos and Mesoscopic Systems' contributed to this
work. We thank Alex Dragt for several stimulating conversations.

\appendix 

\section{$1\leq k\epsilon ^{3/2},k\epsilon ^2<<1$}

We here indicate how to find the resonant solution if $kb^3\simeq 1,kb^2>>1,$
and $kb^4<<1.$ We expand $S_0$ about the nominal stationary point $\Theta
_{pq}$ to third order: 
\begin{eqnarray}
S_0(\theta -\theta ^{\prime })+l_{pq}(\theta ^{\prime }-\theta ) &\simeq
&S_0(\Theta _{pq})+l_{pq}\Theta _{pq}  \nonumber \\
&&-k\frac d4\delta \theta ^2+\frac 18kdd_1\delta \theta ^3.  \label{S03}
\end{eqnarray}
The order $\epsilon $ term is 
\begin{equation}
S_2(\theta ,\theta ^{\prime })\simeq kb^2(dV_2(\theta )+dV_2^{\prime
}(\theta )\delta \theta )  \label{S21}
\end{equation}
and the order $\epsilon ^2$ term is 
\begin{equation}
S_4(\theta ,\theta ^{\prime })=kdb^4V_4(\theta ).  \label{S40}
\end{equation}
We have normalized to the second derivative of $S_0\,$ which is written in
terms of the constant $d.$ Note that $V_2^{\prime }$ is not the derivative
of $V_2.$

The wavefunction is 
\begin{equation}
\psi =\exp (-g)\exp ikd\left( bf_1+b^2f_2+b^3f_3+b^4f_4)\right) .
\label{psi4}
\end{equation}
We expand 
\begin{eqnarray}
f_1(\theta ^{\prime }) &=&f_1(\theta +\Theta _{pq})+\delta \theta
f_1^{\prime }(\theta +\Theta _{pq})+\delta \theta ^2f_1^{\prime \prime
}(\theta +\Theta _{pq})  \nonumber \\
f_2(\theta ^{\prime }) &=&f_2(\theta +\Theta _{pq})+\delta \theta
f_2^{\prime }(\theta +\Theta _{pq})  \nonumber \\
f_3(\theta ^{\prime }) &=&f_3(\theta +\Theta _{pq})  \label{fExpand}
\end{eqnarray}
We know $f_{1\text{ }}$ and the part of $f_2$ which has no Fourier
components which are multiples of $q.$ We investigate the conditions on the
solution which arise at order $kb^3.$

The stationary phase point of the integral satisfies the equation 
\[
-\frac 12\delta \theta +bf_1^{\prime }+24d_1\delta \theta ^2+bf_1^{\prime
\prime }\delta \theta +b^2f_2^{\prime }+b^2V_2^{\prime }=0. 
\]
The last four terms are of order $b^2$, while the first two are of order $b,$
since the leading expression for $\delta \theta $ is of order $b.$ Thus, we
find 
\[
\delta \theta =2bf_1^{\prime }+2b^2(24d_1(f_1^{\prime })^2+2f_1^{\prime
}f_1^{\prime \prime }+f_2^{\prime }+V_2^{\prime }) 
\]
The integration is performed, yielding a condition at order $b^3$%
\begin{eqnarray}
&&d_1f_1^{\prime 3}+2f_1^{\prime 2}f_1^{\prime \prime }+2f_1^{\prime
}(f_2^{\prime }+V_2^{\prime })  \nonumber \\
&=&f_3(\theta )-f_3(\theta +\Theta _{pq})+const  \label{f3}
\end{eqnarray}
Taking the q-average of both sides, we have an equation determining $\bar f%
_2 $, the part of $f_2$ invariant under q-average. 
\begin{equation}
\bar f_{2q}^{\prime }=-\bar V_{2q}^{\prime }-f_1^{\prime }f_1^{\prime \prime
}-\frac 43d_1(f_1^{\prime })^2+\frac c{f_1^{\prime }}  \label{f2bar}
\end{equation}

The part of $f_3$ vanishing under $q$-average is determined from 
\[
f_3(\theta )-f_3(\theta +\Theta _{pq})=2f_1^{\prime }(f_2^{\prime }-\bar f%
_2^{\prime }+V_2^{\prime }-\bar V_{2q}^{\prime }) 
\]
and $\bar f_3$ is determined by the next order condition. Note that if $f_1$
is double valued, so if $f_3.$ It can therefore change the shape of the
resonance islands.

\section{No small parameter}

We add some further remarks concerning the quantization of the bouncing ball
states in the stadium billiard and some related systems.

A WKB quantization condition of Eq.(\ref{fpStad}) is 
\[
2k\int_0^{x+}dx\sqrt{E_m-V(x)}=2ka\sqrt{E_m}+k\frac \pi 2E_m=\pi (m+\nu ). 
\]
The term in $E_m$ comes from the integral in the region $x>a$ and is smaller
than the leading term.

Tanner\cite{Tanner}, by a method using only periodic orbits and with a
different surface of section than we use, finds after an impressive
calculation, an approximate energy equation equivalent to

\[
\sqrt{E_m}=\frac{\pi (m+1/2)}{2ka}.\text{ } 
\]
That is, he takes $\nu =\frac 12$ and neglects the integral for $\left|
x\right| >a.$ He considers just the states odd under reflection $%
x\rightarrow -x$, which for us means $m$ odd, i.e. $m\rightarrow 2m-1,$ $%
E_m=\left( \pi (m-1/4)/ka\right) ^2.$

The equation for the energy of the odd-odd states is then 
\[
k_{n,m}=\pi n+k_{n,m}E_m\approx \pi \sqrt{n^2+a^{-2}(m-\frac 14)^2} 
\]
Tanner finds this to be a good approximation, although he applies it outside
the range of our derivation, that is, for $a=5$, he takes $n=1,2,3$ and $%
m=1..9.$

We find things a bit more complicated, especially in determining the Maslov
exponent. This will be reported separately, however.

\begin{figure}[tbp]
{\hspace*{0.9cm}\psfig{figure=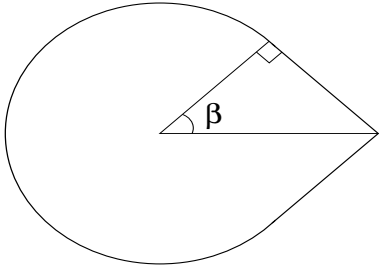,height=5cm,width=3.5cm,angle=270}} {%
\vspace*{.13in}}
\caption[ty]{The ice cream cone billiard with parameter $\beta =\pi /4.$}
\label{fig:10}
\end{figure}

Next, consider an ice cream cone billiard, which is a unit circle for $%
\left| \theta \right| >\beta ,$ and a triangular shaped region for $\left|
\theta \right| <\beta .$ See Fig. 10. Microwave experiments on such a
billiard have been recently carried out\cite{Fromhold}. The equation for the
straight line segments is 
\[
\left| y\right| =\csc \beta -x\cot \beta 
\]
for $\cos \beta <x<\sec \beta .$

Using the unit circle as nominal integrable system, the billiard boundary as
SS, and $\theta $ as the parameter determining position on the SS, we find
for the action, on and near the circular section, 
\[
S(\theta ,\theta ^{\prime })=2k\sin \frac 12(\theta -\theta ^{\prime
})+k(\xi (\theta )+\xi (\theta ^{\prime })) 
\]
where $\xi (\theta )\approx \frac 12(\left| \theta \right| -\beta )^2;$ $%
\left| \theta \right| <\beta $, $=0;$ $\beta <\left| \theta \right| <\pi .$

This leads to essentially the same WKB problem as in the stadium case just
discussed. The change of sign of the second derivative is compensated by the
change of sign of the deviation from the integrable system. Because,
however, zero angular momentum makes $\theta ^{\prime }\approx \theta \pm
\pi $, the effective potential $\bar V_2=\frac 12$ $(\left| \theta \right|
-\beta )^2;\left| \theta \right| <\beta ,$ $\frac 12(\left| \pi -\theta
\right| -\beta )^2,$ $\left| \pi -\theta \right| <\beta ,$ $=0;$ $\beta
<\left| \theta \right| <\pi -\beta .$ In other words, there are two
`potential wells' whose confining walls rise at half the rate of the stadium
case.

The potential formulation seems to hold in the stadium case, even for $%
a\rightarrow 0,$ or in the ice cream cone case, for $\beta \rightarrow \pi .$
In fact, in the latter case, the ice cream cone billiard approaches a very
long stadium. It is a good exercise, however, to show that in these cases,
the solution is such that the implicit small parameter, $E_m$ can no longer
be taken as small enough to justify the procedure.

Of course, it is not difficult to extend these results to deal with cases
having an additional small perturbation. For example, the flat sides of the
stadium might be given some small wriggles. An example of this type that has
been published\cite{stadSlant} makes the radius at one end of the stadium
billiard slightly larger than at the other end, so that the stadium flat
sides are not quite parallel.


\begin{references}
\bibitem{VanVleck}  J. H. Van Vleck, {\em Quantum Principles and Line
Spectra, }Bull. Natl. Res. Council, {\bf 10(54),} 1 (1926).

\bibitem{Gutz}  M. C. Gutzwiller, {\em Chaos in Classical and Quantum
Mechanics, }(Springer, New York, 1991).

\bibitem{LandL}  A. J. Lichtenberg and M. A. Lieberman, {\em Regular and
Chaotic Dynamics, }second ed. (Springer, New York, 1992).

\bibitem{Lazutkin}  V. F. Lazutkin {\em KAM Theory and Semiclassical
Approximations to Eigenfunctions, }(Springer-Verlag, Berlin-Heidelberg,
1993).

\bibitem{bogolss}  E. B. Bogomolny, Nonlinearity {\bf 5, }805 (1992).

\bibitem{dorsmil}  E. Doron and U. Smilansky, Phys. Rev. Lett. {\bf 68},
1255 (1992); B. Dietz and U. Smilansky, Chaos {\bf 3}(4), 581 (1993).

\bibitem{BK}  M. V. Berry and J. P. Keating, J. Phys. {\bf 23, }4839 (1990);
M. V. Berry and J. P. Keating, Proc. Roy. Soc.(London) {\bf A437}, 151-173
(1992).

\bibitem{CvitEck}  P. Cvitanovi\'c and B. Eckhardt, Phys. Rev. Letters {\bf %
63}, 823 (1989).

\bibitem{TLit}  P. A. Boasman, Nonlinearity {\bf 7, }485, (1994); B.
Lauritzen, CHAOS {\bf 2}, 409 (1992); T. Szeredi, J. H. Lefebvre and D. A.
Goodings, Nonlinearity {\bf 7}, 1463 (1994), M. R. Haggerty, Phys. Rev. {\bf %
E52}, 389 (1995).

\bibitem{PranFred}  B. Georgeot and R. E. Prange, Phys. Rev. Lett. {\bf 74},
2851 (1995); B. Georgeot and R. E. Prange, Phys. Rev. Lett. {\bf 74, }4110
(1995); S. Fishman, B. Georgeot and R. E. Prange, J. Phys. {\bf A29},
919-937 (1996); B. Georgeot and R. E. Prange, in {\em Quantum Classical
Correspondence: Proceedings of the 4th Drexel Symposium, }pp 429-444
(International Press, 1997), Eds. d. H. Feng and B. L. Hu.

\bibitem{pnz1}  R. E. Prange, R. Narevich and O. Zaitsev,
chao-dyn/9802019(1998).

\bibitem{Shuryak}  E. V. Shuryak, Sov. Phys. JETP {\bf 44}, 1070 (1976).

\bibitem{transition}  R. Grempel, S. Fishman and R. E. Prange, Phys. Rev.
Letters, {\bf 53, }1212 (1984).

\bibitem{BG}  G. D. Birkhoff, {\em Dynamical Systems }(Am. Math. Soc., New
York, 1966). Vol. IX; F. G. Gustavson, Astron. J. {\bf 71, }670, (1966);

\bibitem{Eckhardt}  B. Eckhardt, J. Phys. A: Math. Gen. {\bf 19}, 2961-2972
(1986).

\bibitem{DD}  B. T. Darling and D. M. Dennison, Phys. Rev. {\bf 57, }128
(1940).

\bibitem{ozo}  A. M. Ozorio de Almeida, J. Phys. Chem. {\bf 88}, 6139 (1984).

\bibitem{Ezra2}  C. C. Martens and G. S. Ezra, J. Chem. Phys. {\bf 83}(6),
2990 (1985). This paper contains a good review in its introduction.

\bibitem{NandM}  D. W. Noid and R. A. Marcus, J. Chem. Phys. {\bf 62}, 2119
(1975).

\bibitem{Ezra1}  G. S. Ezra, to appear in {\em Advances in Classical
Trajectory Methods: Comparisons of classical and quantum dynamics, }W. L.
Hase, ed. (JAI Press, 1998). This paper contains many references to related
work.

\bibitem{Ozorio}  A. M. Ozorio de Almeida, {\em Hamiltonian Systems: Chaos
and Quantization, }(Cambridge University Press, Cambridge, 1988).

\bibitem{OandH}  A. M. Ozorio de Almeida and J. H. Hannay, J. Phys. A:
Math. Gen. {\bf 20,} 5873-5883 (1987).

\bibitem{Ullmo}  D. Ullmo, {\em et al}, Phys. Rev. {\bf E54, }136 (1996); S.
Tomsovic, {\em et al, }Phys. Rev. Lett. {\bf 75}, 4346 (1995).

\bibitem{Stone}  J. D. N\"ockel, {\em et al, }Optics Lett.{\bf 21}, 1609
(1996); J.D. N\"ockel and A. D. Stone, Nature {\bf 385, }45, (1997).

\bibitem{BCL}  F. Borgonovi, {\em et al}, Phys. Rev. Lett. {\bf 77,} 4744
(1996).

\bibitem{FS}  K. M. Frahm and D. M. Shepelyansky, Phys. Rev. Lett. {\bf 78, }%
1440 (1997); ibid. {\bf 79, }1833 (1997).

\bibitem{Vall}  M. Valli\`eres, ITAMP Workshop, Harvard, (May 1998).

\bibitem{MF}  P. M. Morse and H. Feshbach, {\em Methods of Theoretical
Physics, }(McGraw-Hill, New York, 1953).

\bibitem{Borgo}  F. Borgonovi, Phys. Rev. Lett. {\bf 80}, 4653 (1998).

\bibitem{CP}  G. Casati and T. Prosen, preprint, cond-mat 9803340 (1998).

\bibitem{Zait1}  O. Zaitsev, Ph. D. Thesis and to be published.

\bibitem{Cantori}  I. C. Percival, AIP Conf. Proc. {\bf 57}, 301 (1979).

\bibitem{RepC}  S. Aubry, in {\em Solitons and Condensed Matter Physics,}
ed. A. Bishop and T. Schneider (Springer, Berlin, 1978), p.264-278; G.
Radons and R. E. Prange, Phys. Rev. Lett. {\bf 61, }1691 (1988).

\bibitem{Tanner}  G. Tanner, J. Phys. A: Math. Gen. {\bf 30},
2863-2888 (1997).

\bibitem{ABGOP}  R. Bl\"umel, {\em et al, }Physical Review Letters {\bf 76},
2476 (1996); R. E. Prange, {\em et al,} Phys. Rev. {\bf E53, 207, (1996).}

\bibitem{Chirikov}  B. V. Chirikov, Phys. Rep. {\bf 52, }263 (1979).

\bibitem{FGP}  S. Fishman, D. R. Grempel and R. E. Prange, Phys. Rev. Lett. 
{\bf 49}, 509 (1982).

\bibitem{SmAcc}  H. Primack and U. Smilansky, J. Phys. A: Math. Gen. {\bf 31},
6253-6277 (1998).

\bibitem{CATun}  S. Tomsovic and D. Ullmo, Phys. Rev. {\bf E50}, 145 (1994).

\bibitem{Maslov}  V. P. Maslov and M. V. Fedorjuk, {\em Semiclassical
Approximation in Quantum Mechanics, }(Reidel, Boston, 1981).

\bibitem{pseudo} P. J. Richens and M. V. Berry, Physica D {\bf 2}, 495 (1981).

\bibitem{Fromhold}  T. H. Fromhold, Seminar at Newton Institute, December
1997.

\bibitem{stadSlant}  H. Primack and U. Smilansky, J. Phys. A: Math. Gen. 
{\bf 27}, 4439 (1994).
\end{references}
\end{document}